\documentclass[
reprint,
superscriptaddress,
nofootinbib,
amsmath,amssymb,
aps,
prl,
floatfix,
]{revtex4-1}

\usepackage{graphicx}% Include figure files
\usepackage{dcolumn}% Align table columns on decimal point
\usepackage{bm}% bold math
\usepackage{epstopdf}
\usepackage{braket}
\begin{document}

\title{Trapped-Ion Spin-Motion Coupling with Microwaves and a Near-Motional Oscillating Magnetic Field Gradient}

\author{R. Srinivas}
\email{raghavendra.srinivas@colorado.edu}
\affiliation{Time and Frequency Division, National Institute of Standards and Technology, 325 Broadway, Boulder, Colorado 80305, USA}
\affiliation{Department of Physics, University of Colorado, Boulder, Colorado 80309, USA}
\author{S. C. Burd}
\affiliation{Time and Frequency Division, National Institute of Standards and Technology, 325 Broadway, Boulder, Colorado 80305, USA}
\affiliation{Department of Physics, University of Colorado, Boulder, Colorado 80309, USA}
\author{R. T. Sutherland}
\affiliation{Physics Division, Physical and Life Sciences, Lawrence Livermore National Laboratory, Livermore, California 94550, USA}
\author{A. C. Wilson}
\affiliation{Time and Frequency Division, National Institute of Standards and Technology, 325 Broadway, Boulder, Colorado 80305, USA}
\author{\break D. J. Wineland}
\affiliation{Time and Frequency Division, National Institute of Standards and Technology, 325 Broadway, Boulder, Colorado 80305, USA}
\affiliation{Department of Physics, University of Colorado, Boulder, Colorado 80309, USA}
\affiliation{Department of Physics, University of Oregon, Eugene, Oregon 97403, USA}
\author{D. Leibfried}
\affiliation{Time and Frequency Division, National Institute of Standards and Technology, 325 Broadway, Boulder, Colorado 80305, USA}
\author{D. T. C. Allcock}
\affiliation{Time and Frequency Division, National Institute of Standards and Technology, 325 Broadway, Boulder, Colorado 80305, USA}
\affiliation{Department of Physics, University of Colorado, Boulder, Colorado 80309, USA}
\affiliation{Department of Physics, University of Oregon, Eugene, Oregon 97403, USA}
\author{D. H. Slichter}
\email{daniel.slichter@nist.gov}
\affiliation{Time and Frequency Division, National Institute of Standards and Technology, 325 Broadway, Boulder, Colorado 80305, USA}

\date{\today}

\begin{abstract}
We present a new method of spin-motion coupling for trapped ions using microwaves and a magnetic field gradient oscillating close to the ions' motional frequency. We demonstrate and characterize this coupling experimentally using a single ion in a surface-electrode trap that incorporates current-carrying electrodes to generate the microwave field and the oscillating magnetic field gradient. Using this method, we perform resolved-sideband cooling of a single motional mode to its ground state.
\end{abstract}

%\pacs{Valid PACS appear here}% PACS, the Physics and Astronomy Classification Scheme.

%\keywords{Suggested keywords}%Use showkeys class option if keyword display desired

\maketitle

Coupling the internal spin states of trapped ions to their motion is essential for applications in quantum information processing, quantum simulation, and metrology~\cite{Wineland1998, Blatt2008, Blatt2012}, such as quantum logic gates~\cite{Cirac1995, Monroe1995}, simulations of many-body spin systems~\cite{Kim2010, Lanyon2011, Britton2012}, and quantum logic spectroscopy for optical clocks~\cite{Schmidt2005} and molecules~\cite{Wolf2016}. This coupling requires a field gradient across the ions' wave function, and is usually accomplished using laser-induced interactions. However, photon scattering errors are a fundamental limit for laser-based spin-motion coupling~\cite{Ozeri2007}, and are the leading error in the highest-fidelity quantum logic gates~\cite{Ballance2016, Gaebler2016}. Laser-induced coupling is also sensitive to fluctuations in the optical phase and intensity at the ion, which can be experimentally demanding to mitigate. Alternative laser-free methods, which are not limited by photon scattering and offer improved phase and amplitude stability, use microwave radiation and magnetic field gradients to perform the spin-motion coupling. These fields and gradients can be generated using current-carrying wires integrated in a surface-electrode trap~\cite{Seidelin2006}, a promising platform for large scale quantum computing or simulation with ions. The integrated microwave circuitry can also be used to perform high-fidelity single-qubit gates~\cite{Brown2011, Harty2014} and individual ion addressing~\cite{Warring2013a, Craik2014}.

Laser-free spin-motion coupling has been proposed and demonstrated using either a static magnetic field gradient with separate microwave fields~\cite{Mintert2001, Johanning2009, Khromova2012} or oscillating magnetic field gradients close to the qubit frequency~\cite{Ospelkaus2008, Ospelkaus2011}. High-fidelity two-qubit gates have been performed with these methods~\cite{Weidt2016, Harty2016}. However, the near-qubit oscillating gradient method requires large currents near the qubit frequency to generate a strong gradient, which can be technically challenging for typical gigahertz-frequency hyperfine qubits. Furthermore, most entangling gates require multiple such currents at different frequencies in the same trap electrodes. In contrast, the static gradient scheme enables multiple interactions to be generated using only a single strong gradient with multiple weak microwave fields. However, the spin-motion coupling strength for this scheme decreases rapidly with increasing motional frequency for a given microwave current. Higher motional frequencies are desirable to mitigate the effects of anomalous heating~\cite{Brownutt2015} and to reduce the average phonon occupation of the motional modes after Doppler cooling. This reduces the time for ground-state cooling, a requirement for many quantum information and simulation experiments.

In this work, we demonstrate a new technique for spin-motion coupling in trapped ions using microwaves and a near-field magnetic field gradient oscillating close to the ions' motional frequency. This method is a generalization of the static-gradient scheme, and enables stronger spin-motion coupling for a given motional frequency, gradient strength, and microwave field amplitude. In particular, strong spin-motion coupling with low microwave power can be maintained even at high motional frequencies.  This near-motional gradient can be more efficiently generated than an equivalent gigahertz-frequency gradient, and only one strong gradient is required to implement multiple simultaneous spin-motion coupling interactions.  Here, we characterize the spin-motion coupling strength by tuning the microwave and gradient parameters and identify optimal working regimes.  We cool an ion to its motional ground state as a proof of principle.

\begin{figure}[tb]
\includegraphics[width=0.5\textwidth]{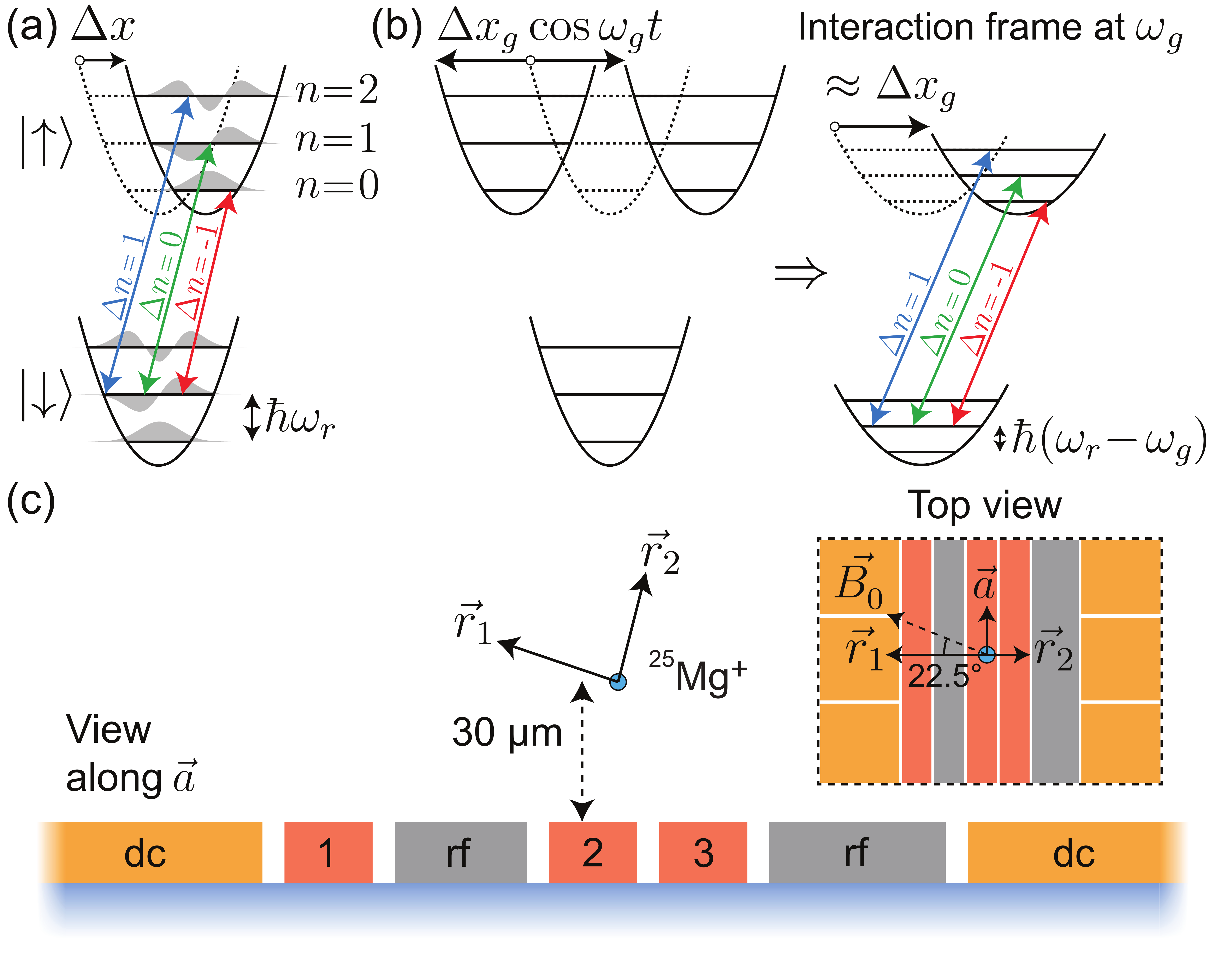}
\centering
\caption{\label{general} Schematic description of a qubit coupled to a harmonic oscillator with a spin-dependent displacement from (a) a static gradient or (b) an oscillating gradient.  An additional microwave field drives $\ket{\downarrow}\leftrightarrow\ket{\uparrow}$ transitions.  (a) For a static gradient, detuning the microwave field by $\pm\,\omega_r$ drives sideband transitions with $\Delta n = \pm1$. (b) An oscillating gradient at $\omega_g$ is formally equivalent to a static gradient in the interaction frame oscillating at $\omega_g$, ignoring fast-oscillating terms.  The sideband transitions now occur at detunings $\pm (\omega_r-\omega_g)$. (c) Schematic of the surface electrode trap. The ion has three motional modes, $\vec{a}$ along the trap axis and $\vec{r}_1$ and $\vec{r}_2$ perpendicular to the trap axis. The dc and rf electrodes provide trapping potentials, while oscillating currents in electrodes 1, 2, and 3 generate magnetic fields and magnetic field gradients at the ion. A static magnetic field $\vec{B}_0$ parallel to the plane of the trap defines the quantization axis $\hat{z}$.}
\end{figure}

The physics underlying this spin-motion coupling involves a magnetic field gradient which causes a spin-dependent spatial displacement of the ion. We consider the case of a microwave-frequency hyperfine qubit, but in general this would apply to any two-level system, as long as the gradient displaces one of the qubit states relative to the other. The basic physics has been observed previously using the intensity gradient from a running optical lattice~\cite{Ding2014}, but with the aforementioned limitations of a laser-based approach. The harmonically confined ion has internal ``spin" states labeled $\ket{\downarrow}$ and $\ket{\uparrow}$, separated in energy by $\hbar\omega_0$, as well as motional states $\ket{n}$ separated by $\hbar \omega_r$. In the absence of any spin-dependent displacement, a microwave field driving the $\ket{\downarrow} \leftrightarrow \ket{\uparrow}$ transition can only drive spin flips that leave the ion's motional state unchanged ($\Delta n = 0$). However, if a gradient displaces $\ket{\uparrow}$ relative to $\ket{\downarrow}$ (see Fig.~\ref{general}(a)), the overlap between their corresponding motional state wave functions is modified, enabling a microwave field with detuning $\delta = \pm \omega_r$ from the qubit frequency $\omega_0$ to drive motion-changing sideband transitions ($\Delta n = \pm1)$; this can be viewed as a change in the Franck-Condon factors~\cite{Foerster2009, Hu2011}. For small displacements, the sideband Rabi frequency increases with the magnitude of the displacement $\Delta x$, and for a static gradient the explicit dependence of the sideband Rabi frequency on the motional frequency is $\Omega_{\text{sb}}\propto \omega_r^{-3/2}$. If the gradient is instead oscillating at a frequency $\omega_g$, the spin-dependent displacement $\Delta x_g$ can be larger for a given gradient strength if the ion motion is being driven closer to resonance~\cite{Welzel2011}, as shown in Fig.~\ref{general}(b). Sideband transitions now occur at $\delta = \pm(\omega_r-\omega_g)$, made apparent by transforming into an interaction frame oscillating at $\omega_g$ and making a rotating wave approximation.  The gradient in this frame appears static, with a modified ``motional frequency'' $(\omega_r-\omega_g)$. The sideband Rabi frequency scales as ${\Omega_{\text{sb}}\propto[\sqrt{\omega_r}(\omega_r-\omega_g)]^{-1}}$, which grows larger as $\omega_g\rightarrow\omega_r$, and reduces to the static gradient case when $\omega_g=0$.  Thus the spin-motion coupling strength can be larger than the static gradient case for a given $\omega_r$, gradient strength, and microwave field amplitude.  An additional, weaker set of sidebands (highlighted by transforming into an interaction frame oscillating at $-\omega_g$ instead) appears at $\delta = \pm(\omega_r+\omega_g)$ with $\Omega_{\text{sb}} \propto  [\sqrt{\omega_r}(\omega_r+\omega_g)]^{-1}$.

The system Hamiltonian $\hat{H}(t)$ with fields oscillating at $\omega_g$ and $\omega_0+\delta$ can be written in the interaction picture with respect to $\hat{H}_0 = \frac{\hbar \omega_0}{2}\hat{\sigma}_z + \hbar\omega_g \hat{a}^\dagger \hat{a}$ as
\begin{align}
\label{Hfull}
\begin{split}
\hat{H}_I(t)= &\hbar\Omega_g \hat{\sigma}_z\left[(\hat{a}+\hat{a}^\dagger) + (\hat{a}e^{-2i\omega_gt}+\hat{a}^\dagger e^{2i\omega_gt})\right]
\\&+2\hbar\Omega_z\cos(\omega_gt)\hat{\sigma}_z +\hbar(\omega_r-\omega_g)\hat{a}^\dagger\hat{a}
\\
&+\hbar\Omega_\mu\left(\hat{\sigma}_+e^{-i\delta t}
+\hat{\sigma}_-e^{i\delta t}\right),
\end{split}
\end{align}
\noindent where $\hat{\sigma}_z=\ket{\uparrow}\bra{\uparrow}-\ket{\downarrow}\bra{\downarrow}$, $\hat{\sigma}_+ = \ket{\uparrow}\bra{\downarrow}$, $\hat{\sigma}_- = \ket{\downarrow}\bra{\uparrow}$, and $\hat{a}^{\dagger}$ and $\hat{a}$ are creation and annihilation operators for the ion motion. The coupling of the qubit states by the microwave field is characterized by $\Omega_\mu$~\cite{supplementary}.  The spin and motion are coupled (with strength $\propto\Omega_g$) by the gradient of a magnetic field $B_g$ along the quantization axis oscillating at $\omega_g$. In this interaction frame, ignoring faster terms at $2\omega_g$, the gradient appears static with a modified ``motional frequency" $\omega_r-\omega_g$. In general, $B_g$ is nonzero at the ion position, giving an additional term in the Hamiltonian proportional to $\Omega_z$.  We define $\Omega_g$ and $\Omega_z$ as

\begin{align}
\label{eq_omegas}
\begin{split}
\Omega_g &\equiv \frac{r_0(\hat{r}\cdot\nabla B_g)}{4}\frac{d\omega_0}{dB_z}\Big|_{B_z=|\vec{B}_0|} \\
\Omega_z &\equiv \frac{B_g}{4}\frac{d\omega_0}{dB_z}\Big|_{B_z=|\vec{B}_0|}.
\end{split}
\end{align}
\noindent The sensitivity of the qubit frequency $\omega_0$ to changes in the magnetic field $B_z$ along the quantization axis $\hat{z}$ (defined by a static magnetic field $\vec{B}_0$) is described by $d\omega_0/dB_z$, and $r_0=\sqrt{\hbar/2M\omega_r}$ is the ground-state extent of the ion wave function for the motional mode along the $\hat{r}$ direction for ion mass $M$. This gives an implicit $\omega_r^{-1/2}$ dependence to $\Omega_g$.  

\begin{figure*}[tb]
\includegraphics[width=\textwidth]{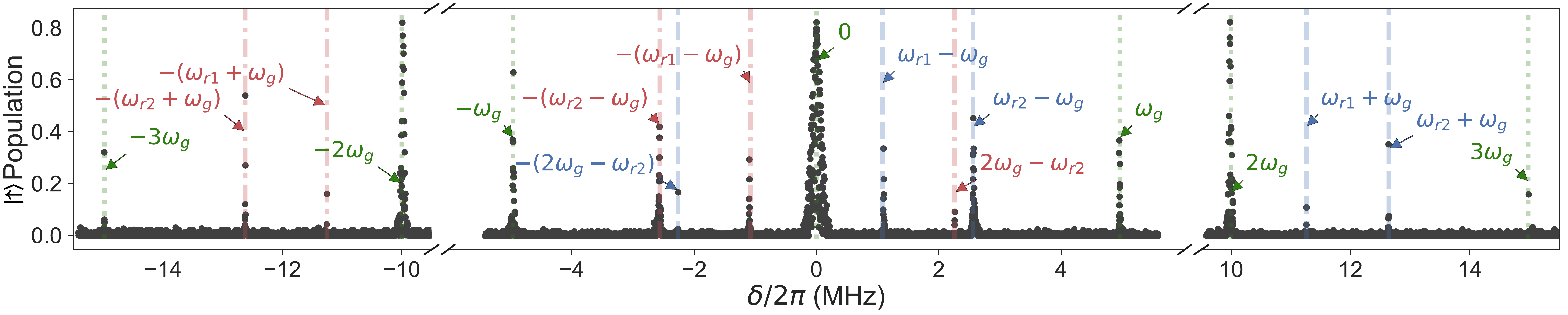}
\centering
\caption{\label{lf_spectrum} Microwave spectroscopy in the presence of a magnetic field with a gradient oscillating at $\omega_g/2\pi = 5$\,MHz. Trap radial frequencies are ${(\omega_{r_1},\omega_{r_2})/2\pi \approx (6.2, 7.6)}$\,MHz. An additional microwave pulse with detuning $\delta$ from the qubit frequency is applied for 500\,$\mu$s. The ion is initialized close to the Doppler temperature in the $\ket{\downarrow}$ state and the population in the $\ket{\uparrow}$ state is measured at the end of the pulse. Spin-flip transitions with $\Delta n=0$ (green dotted lines) occur when $\delta = \pm\,m\omega_g$, and blue (red) motional sideband transitions with $\Delta n=1$($\Delta n=-1$) [blue dashed (red dash-dotted) lines] occur when $\delta = +(\omega_{ri} \pm \omega_g)$ [$\delta = -(\omega_{ri} \pm \omega_g$)]. Weak sideband transitions at $\pm(\omega_{r2}-2\omega_g)$ correspond to higher-order interactions~\cite{supplementary}.  Population error bars are omitted for clarity.}
\end{figure*}

By transforming Eq.~(\ref{Hfull}) into the interaction picture with respect to the terms in the first two lines~\cite{supplementary}, different interactions are obtained for specific values of $\delta$.  Sideband transitions occur at $\delta = \pm (\omega_r - \omega_g)$ or $\delta = \pm (\omega_r + \omega_g)$.  For the first case, the Hamiltonian is given by
\begin{align}
\label{eq_sb}
\hat{H}_{\text{sb}} =  \pm&\hbar\Omega_{\text{sb}}(\hat{\sigma}_\pm \hat{a}^{\dagger}+\hat{\sigma}_\mp \hat{a}),
\end{align}
\noindent where the upper (lower) sign choice corresponds to the blue (red) sideband interaction. The sideband Rabi frequency is ${\Omega_{\text{sb}}\equiv\frac{2\Omega_g\Omega_\mu}{\omega_r-\omega_g}J_0(\frac{4\Omega_z}{\omega_g})}$, where $J_0$ is the zeroth-order Bessel function of the first kind. For the case $\delta = \omega_r + \omega_g$, the denominator $\omega_r-\omega_g$ in $\Omega_{\text{sb}}$ is replaced by $\omega_r+\omega_g$.  The Rabi frequency $\Omega_{\text{sb}}$ is maximized when the argument of ${J_0(\frac{4\Omega_z}{\omega_g})}$ is zero, which corresponds to $B_g=0$ (importantly, this does not imply $\nabla B_g=0$). When $B_g$ is nonzero, the qubit frequency is modulated, enabling spin-flip transitions that do not change the motional state of the ion. These transitions occur when $\delta=m\omega_g$ for integer $m$, and the resulting Hamiltonian is
\begin{align}
\label{eq_carrier}
\hat{H}_m &= \hbar\Omega_m\hat{\sigma}_x,
\end{align}
\noindent where $\Omega_m\equiv\Omega_\mu J_m(\frac{4\Omega_z}
{\omega_g})$ is the spin-flip Rabi frequency, and $J_m$ is the $m^\mathrm{th}$-order Bessel function of the first kind. A similar effect has been observed from residual magnetic fields generated by the rf trapping potentials~\cite{Meir2018}.

In our experiment, the gradient and microwave fields are generated by currents driven through electroplated gold electrodes of a cryogenic (18\,K) linear surface-electrode Paul trap, labeled as electrodes 1, 2, and 3 in Fig.~\ref{general}(c). A single $^{25}$Mg$^+$ ion is trapped ${\approx\,30\,\mu}$m above the surface with motional frequencies $(\omega_a, \omega_{r_1},\omega_{r_2})/2\pi \approx (3.2, 6.2, 7.6)$\,MHz, where $\vec{a}$ is along the axis of the trap, and $\vec{r}_1$, $\vec{r}_2$ lie in the radial plane. We use the $\ket{F=3,m_F=3}\equiv\ket{\downarrow}$ and $\ket{F=2, m_F=2}\equiv\ket{\uparrow}$ states within the $3^2S_{1/2}$ hyperfine manifold as our qubit, which has a transition frequency of ${\omega_0/2\pi = 1.326}$\,GHz in the externally applied magnetic field $|\vec{B}_0|=21.3$\,mT. The magnetic field sensitivity of this transition is ${(d\omega_0/{dB_z})/2\pi=-19.7}$\,MHz/mT and the magnitude of the static field gives $\sim$100\,MHz of spectral separation between adjacent Zeeman sublevels. We prepare $\ket{\downarrow}$ by optical pumping on the $3^2S_{1/2}\leftrightarrow3^2P_{3/2}$ transition at 280\,nm with $\sigma^+$ polarized light. The qubit is read out by detecting fluorescence from the laser-driven $\ket{\downarrow}\leftrightarrow\ket{3^2P_{3/2}, F=4, m_F=4}$ cycling transition. Before detection, microwave pulses are used to shelve $\ket{\uparrow}$ to the far-detuned $\ket{3^2S_{1/2}, F=2, m_F=-1}$ state. To realize spin-motion coupling, we apply simultaneous currents (which are ramped on and off over 10 $\mu$s) to the trap electrodes at two frequencies, $\omega_g$ and $\omega_0 + \delta$. We apply up to 0.5(1)\,A rms per electrode at $\omega_g/2\pi=5$\,MHz, corresponding to 6(1)\,mW of dissipation in the trap; dissipation from the drive at ${\omega_0+\delta}$ is ${\ll 1\,\mathrm{mW}}$~\cite{supplementary}.

\begin{figure}[tb]
\includegraphics[width=0.485\textwidth]{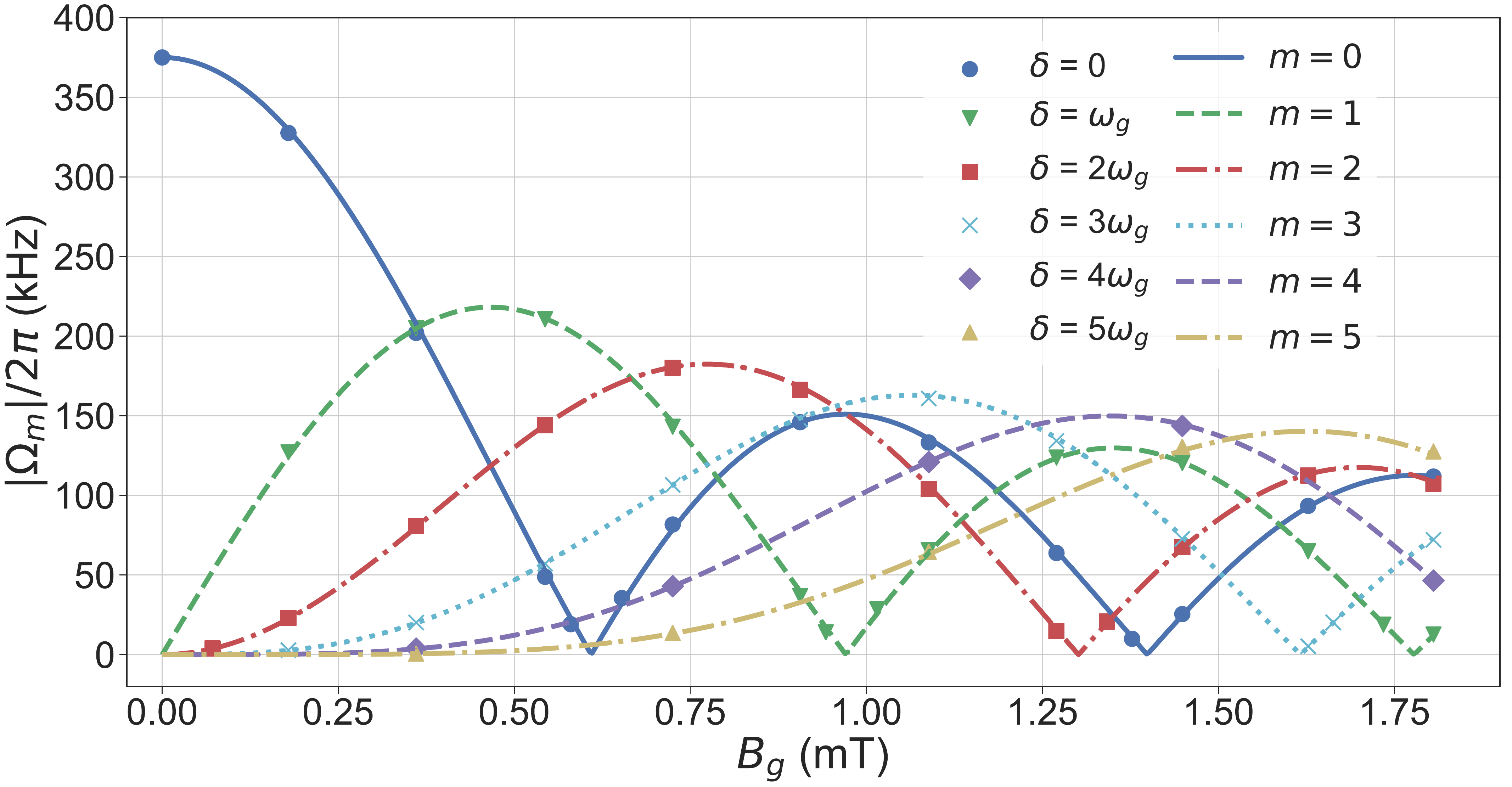}
\centering
\caption{\label{bessel} Dependence of the spin-flip Rabi frequency $\Omega_m$ on $B_g$. Here, we apply a current of variable amplitude oscillating at $\omega_g$ to electrode 2 in addition to a microwave field with detuning $\delta$ and Rabi frequency $\Omega_\mu/2\pi=$ 375\,kHz. Spin-flip transitions occur when $\delta = m\omega_g$ with $\Omega_m = \Omega_\mu J_m(\frac{4\Omega_z}{\omega_g})$, where $\Omega_z\propto B_g$. We calibrate the horizontal axis scale by fitting the data for $\delta=0$ to ${|J_0(\frac{4\Omega_z}{\omega_g})|}$.  The theory curves (dashed) for $m>0$ have no adjustable parameters. Error bars are smaller than the data points.}
\end{figure}

With the oscillating gradient applied, we perform microwave spectroscopy by varying $\delta$ as shown in Fig.~\ref{lf_spectrum}. Spin-flip transitions occur at $\delta = m\omega_g$ since $B_g\ne0$ for this experiment, and motional sideband transitions appear at $\delta = \pm(\omega_{ri}-\omega_g)$ and $\delta = \pm(\omega_{ri} + \omega_g)$, where $\omega_{ri}$ is the frequency of the radial mode $\vec{r}_1$ or $\vec{r}_2$. Weak higher-order sidebands are also visible~\cite{supplementary}.  We do not see sideband transitions for the axial mode $\vec{a}$ as the gradient along the trap axis is (by design) very small. Following Eq.~(\ref{eq_carrier}), the spin-flip transitions are characterized by measuring the Rabi frequencies $\Omega_m$ as a function of $B_g$. The values of $\Omega_m$ versus $B_g$ are shown for $m=\{0,1,2,3,4,5\}$ in Fig.~\ref{bessel}.

Setting $B_g$, and hence $\Omega_z$, to zero maximizes $\Omega_{\text{sb}}$ for the sideband transitions described in Eq.~(\ref{eq_sb}); at this point, $\Omega_{\text{sb}}$ is insensitive to variations in $B_g$ to first order. We set $B_g=0$ by adjusting the relative phases and amplitudes of currents oscillating at $\omega_g$ in electrodes 1, 2, and 3 to minimize $\Omega_{m=1}$, which is proportional to $B_g$ as 
$B_g\rightarrow0$. The magnitude of $\Omega_{\text{sb}}$ is insensitive to components of the oscillating magnetic field perpendicular to $\vec{B}_0$, which produce an ac Zeeman shift on the qubit of less than 100 kHz.  

\begin{figure}[tb]
\includegraphics[width=0.485\textwidth]{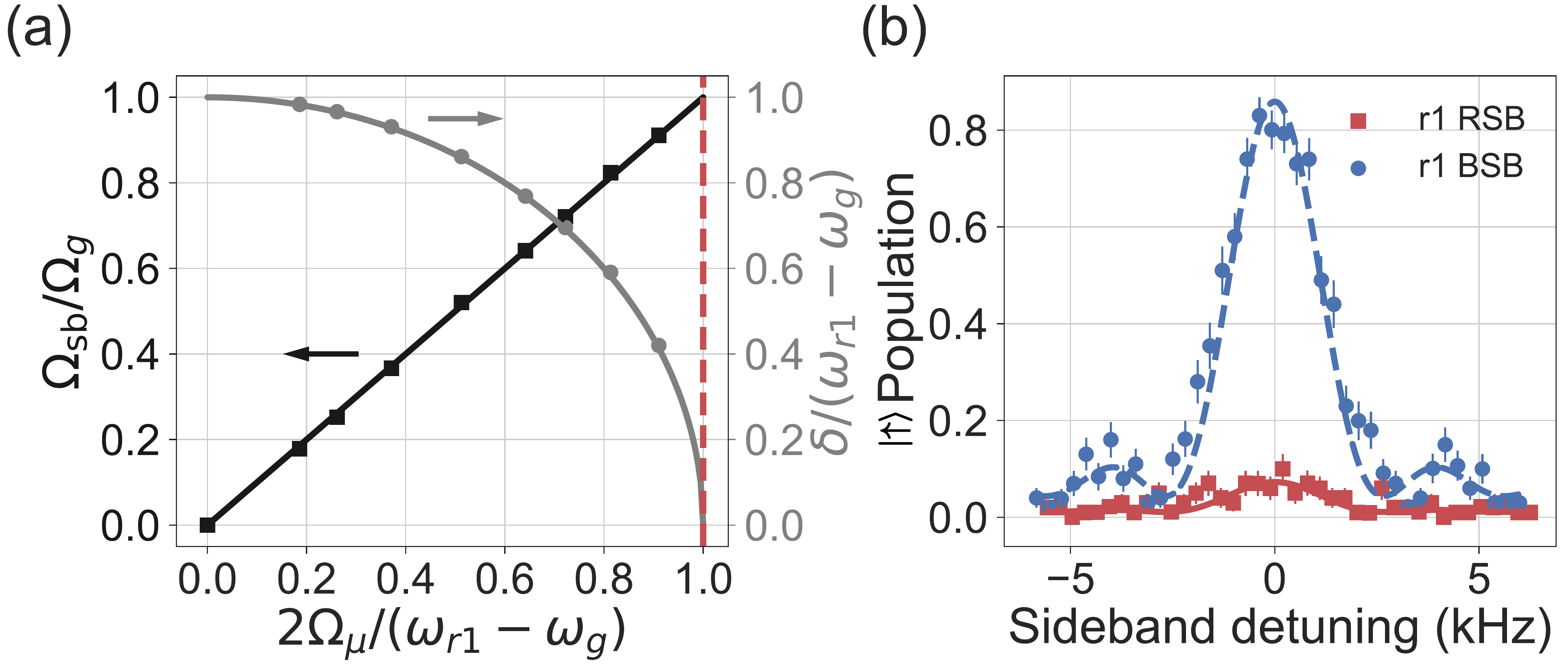}
\centering
\caption{\label{sb_rabi_frequency} Sideband characterization. (a) The normalized blue sideband Rabi frequency ${\Omega_{\text{sb}}/\Omega_g}$ (black squares, left axis) and the normalized microwave detuning $\delta/(\omega_{r1}-\omega_g)$ from Eq.~(\ref{delta_ac}) (gray circles, right axis) are plotted as a function of $2\Omega_\mu/(\omega_{r1}-\omega_g)$.  Error bars are smaller than the data points.  The red dashed line denotes the limit on $\Omega_\mu$ described in the text.  The black line is a linear fit to the data; the gray line is a theoretical plot of Eq.~(\ref{delta_ac}).  
(b) Populations in $\ket{\uparrow}$ after blue (BSB, circles) and red (RSB, squares) sideband analysis pulses on a ground-state-cooled ion in $\ket{\downarrow}$ versus the detuning of the microwaves from the $r_1$ sideband transition.  Both the cooling and the analysis pulses are performed using the oscillating gradient sidebands described in the text. Lines are fits giving $\bar{n}=0.09(7)$.}
\end{figure}

The sideband Rabi frequency also depends on the microwave Rabi frequency $\Omega_\mu$ with $\Omega_{\text{sb}}\propto2\Omega_\mu/(\omega_r-\omega_g)$.  Because of the ac Zeeman shift from the microwave~\cite{supplementary}, the detunings required for the sidebands are shifted from $\delta = \pm( \omega_{r} - \omega_g)$ to
\begin{align}
\label{delta_ac}
 \delta \rightarrow \pm \sqrt{(\omega_{r}-\omega_g)^2-4\Omega_\mu^2}.   
\end{align}
\noindent Thus, as $2\Omega_\mu$ approaches $\omega_r-\omega_g$, $\delta$ approaches zero, causing the microwave field required for the sideband to drive spin flips resonantly. This sets a limit for the maximum usable $\Omega_{\mu}$ for driving sidebands.  Operating close to this limit requires careful microwave pulse shaping to minimize off-resonant qubit excitation. We experimentally verify the relationship of both $\Omega_{\text{sb}}$ and the ac Zeeman shift to $\Omega_\mu$ as shown in Fig.~\ref{sb_rabi_frequency}(a). We determine $\Omega_{\text{sb}}$ by cooling the $r_1$ mode to its ground state and driving the blue sideband transition.  We set ${(\omega_{r1}-\omega_g)/2\pi \approx 1.2}$ MHz and vary $2\Omega_{\mu}/2\pi$ from 0 to 1.1\,MHz.  A linear fit to the data yields $\Omega_g/2\pi =$1.383(6)\,kHz, corresponding to a magnetic field gradient of 49.4(2)\,T/m along $\vec{r}_1$, in agreement with simulations~\cite{supplementary}. 

We can use the resolved red sideband at ${\delta \approx -(\omega_{r1}-\omega_g)}$ to cool the $r_1$ mode, as shown in Fig.~\ref{sb_rabi_frequency}(b).  Starting from a Doppler-cooled mean phonon occupation of $\bar{n}\approx2$, we use a sequence of twelve 150\,$\mu$s pulses with interleaved optical repumping to reach $\bar{n}=0.09(7)$. The total cooling duration is ${\approx 2.5}$ ms, more than an order of magnitude faster than previous demonstrations of microwave cooling using the static gradient scheme~\cite{Weidt2015, Sriarunothai2018}. This speed-up is in part from using higher motional frequencies, which results in a lower initial thermal occupation after Doppler cooling.

For these experiments, we used a Blackman envelope~\cite{Harris1978} to adiabatically ramp the microwave pulse on and off over 10\,$\mu$s. This pulse shaping allows us to operate with $2\Omega_\mu/(\omega_r-\omega_g)=0.9$, and $\Omega_{\text{sb}}$ close to its maximum value of $\Omega_g$. Compared to the static gradient scheme, our scheme allows $\Omega_{\text{sb}}=\Omega_{g}$ to be obtained for $2\Omega_\mu = \omega_r-\omega_g$ instead of a larger $2\Omega_\mu =\omega_r/2$.  Thus, our scheme allows $\Omega_{\text{sb}}$ to be maximized for lower microwave currents.  Larger $\Omega_g$, and thus larger $\Omega_{\text{sb}}$, can be achieved by increasing the current at $\omega_g$ in the trap electrodes.  The maximum current will likely be limited by Joule heating in the trap electrodes.  This heating is significantly lower for a current at megahertz frequencies than at gigahertz frequencies due to the larger skin depth. Furthermore, the magnitude of induced return currents in neighboring electrodes~\cite{Warring2013b} is reduced for lower frequencies, yielding a larger gradient for a given applied current.

By moving $\omega_g$ closer to the motional frequency, we can maintain a given $\Omega_{\text{sb}}$ with lower microwave drive strength $\Omega_{\mu}$.  This produces a smaller ac Zeeman shift, reducing decoherence due to fluctuations in $\Omega_\mu$. Reducing $\omega_r-\omega_g$ also increases the strength of higher-order sidebands that are proportional to $(\frac{\Omega_g}{\omega_r-\omega_g})^n$~\cite{Foerster2009}.  However, the finite impedance of the current-carrying electrodes gives rise to an oscillating potential on these electrodes at $\omega_g$; the resulting electric field drives the ion's motion to an amplitude $\propto(\omega_r^2-\omega_g^2)^{-1}$.  Large-amplitude motion in surface-electrode traps samples increasingly anharmonic regions of the confining potential, and in extreme cases may cause the ion to leave the trap, setting a practical lower limit on the detuning of $\omega_g$ from $\omega_r$.  For the maximum measured $\Omega_{g}$ with $B_g$ nulled at the ion, we measure the amplitude of this electric field to be $\approx 10$\,V/m~\cite{Warring2013b, supplementary}.  The effects of electric fields at $\omega_g$ can be reduced for multiple ions by choosing differential motional modes which are not excited by a uniform electric field.  Alternatively, one could directly compensate these oscillating electric fields using additional electrodes, or use trap designs which place the current-carrying electrodes beneath a metal layer~\cite{Welzel2018} which shields electric fields more strongly than magnetic fields and their gradients.  

In conclusion, we have demonstrated a new method for spin-motion coupling in trapped ions using microwaves and an oscillating magnetic field gradient with a frequency near that of a motional mode.  This and other laser-free techniques for spin-motion coupling eliminate the photon scattering errors inherent in laser-based schemes. Moreover, our method addresses several technical limitations on the implementation of previous laser-free schemes.  We demonstrate this method in a surface-electrode trap, where all control fields are generated using trap-integrated electrodes. Our scheme enables multiple sidebands to be produced using a single strong gradient and one weak microwave field for each sideband.  The multiple sidebands required for M\o lmer-S\o rensen-type two-qubit gates, robust polychromatic two-qubit gates~\cite{Shapira2018, Webb2018}, or mixed-species operations~\cite{Tan2015} can then be implemented simply by adding relatively weak microwave fields. Recent theoretical investigations show that our technique is well suited for new dynamical decoupling schemes that reduce the complexity of microwave quantum logic gates in trapped ions, and that high-fidelity two-qubit gates should be achievable with realistic experimental parameters~\cite{Sutherland2019}.

\begin{acknowledgments}
We thank T. R. Tan and S. C. Webster for helpful discussions, and M. Affolter and C.-W. Chou for comments on the manuscript. These experiments were performed using the ARTIQ control system. R.S., S.C.B., and D.T.C.A. are supported as Associates in the Professional Research Experience Program (PREP) operated jointly by NIST and the University of Colorado Boulder under Award No. 70NANB18H006 from the U.S. Department of Commerce, National Institute of Standards and Technology. This work was supported by ARO, ONR, and the NIST Quantum Information Program. Part of this work was performed under the auspices of the U.S. Department of Energy by Lawrence Livermore National Laboratory under Contract No. DE-AC52-07NA27344 with release number LLNL-JRNL-764019. This Letter is a contribution of NIST, not subject to U.S. copyright.
\end{acknowledgments}

\bibliographystyle{apsrev4-1}
\bibliography{lf_single_ion}

\begin{thebibliography}{41}%
\makeatletter
\providecommand \@ifxundefined [1]{%
 \@ifx{#1\undefined}
}%
\providecommand \@ifnum [1]{%
 \ifnum #1\expandafter \@firstoftwo
 \else \expandafter \@secondoftwo
 \fi
}%
\providecommand \@ifx [1]{%
 \ifx #1\expandafter \@firstoftwo
 \else \expandafter \@secondoftwo
 \fi
}%
\providecommand \natexlab [1]{#1}%
\providecommand \enquote  [1]{``#1''}%
\providecommand \bibnamefont  [1]{#1}%
\providecommand \bibfnamefont [1]{#1}%
\providecommand \citenamefont [1]{#1}%
\providecommand \href@noop [0]{\@secondoftwo}%
\providecommand \href [0]{\begingroup \@sanitize@url \@href}%
\providecommand \@href[1]{\@@startlink{#1}\@@href}%
\providecommand \@@href[1]{\endgroup#1\@@endlink}%
\providecommand \@sanitize@url [0]{\catcode `\\12\catcode `\$12\catcode
  `\&12\catcode `\#12\catcode `\^12\catcode `\_12\catcode `\%12\relax}%
\providecommand \@@startlink[1]{}%
\providecommand \@@endlink[0]{}%
\providecommand \url  [0]{\begingroup\@sanitize@url \@url }%
\providecommand \@url [1]{\endgroup\@href {#1}{\urlprefix }}%
\providecommand \urlprefix  [0]{URL }%
\providecommand \Eprint [0]{\href }%
\providecommand \doibase [0]{http://dx.doi.org/}%
\providecommand \selectlanguage [0]{\@gobble}%
\providecommand \bibinfo  [0]{\@secondoftwo}%
\providecommand \bibfield  [0]{\@secondoftwo}%
\providecommand \translation [1]{[#1]}%
\providecommand \BibitemOpen [0]{}%
\providecommand \bibitemStop [0]{}%
\providecommand \bibitemNoStop [0]{.\EOS\space}%
\providecommand \EOS [0]{\spacefactor3000\relax}%
\providecommand \BibitemShut  [1]{\csname bibitem#1\endcsname}%
\let\auto@bib@innerbib\@empty
%</preamble>
\bibitem [{\citenamefont {Wineland}\ \emph {et~al.}(1998)\citenamefont
  {Wineland}, \citenamefont {Monroe}, \citenamefont {Itano}, \citenamefont
  {Leibfried}, \citenamefont {King},\ and\ \citenamefont
  {Meekhof}}]{Wineland1998}%
  \BibitemOpen
  \bibfield  {author} {\bibinfo {author} {\bibfnamefont {D.~J.}\ \bibnamefont
  {Wineland}}, \bibinfo {author} {\bibfnamefont {C.}~\bibnamefont {Monroe}},
  \bibinfo {author} {\bibfnamefont {W.~M.}\ \bibnamefont {Itano}}, \bibinfo
  {author} {\bibfnamefont {D.}~\bibnamefont {Leibfried}}, \bibinfo {author}
  {\bibfnamefont {B.~E.}\ \bibnamefont {King}}, \ and\ \bibinfo {author}
  {\bibfnamefont {D.~M.}\ \bibnamefont {Meekhof}},\ }\href {\doibase
  10.6028/jres.103.019} {\bibfield  {journal} {\bibinfo  {journal} {J. Res.
  Natl. Inst. Stand. Technol.}\ }\textbf {\bibinfo {volume} {103}},\ \bibinfo
  {pages} {259} (\bibinfo {year} {1998})}\BibitemShut {NoStop}%
\bibitem [{\citenamefont {Blatt}\ and\ \citenamefont
  {Wineland}(2008)}]{Blatt2008}%
  \BibitemOpen
  \bibfield  {author} {\bibinfo {author} {\bibfnamefont {R.}~\bibnamefont
  {Blatt}}\ and\ \bibinfo {author} {\bibfnamefont {D.}~\bibnamefont
  {Wineland}},\ }\href {\doibase 10.1038/nature07125} {\bibfield  {journal}
  {\bibinfo  {journal} {Nature (London)}\ }\textbf {\bibinfo {volume} {453}},\
  \bibinfo {pages} {1008} (\bibinfo {year} {2008})}\BibitemShut {NoStop}%
\bibitem [{\citenamefont {Blatt}\ and\ \citenamefont {Roos}(2012)}]{Blatt2012}%
  \BibitemOpen
  \bibfield  {author} {\bibinfo {author} {\bibfnamefont {R.}~\bibnamefont
  {Blatt}}\ and\ \bibinfo {author} {\bibfnamefont {C.~F.}\ \bibnamefont
  {Roos}},\ }\href {\doibase 10.1038/nphys2252} {\bibfield  {journal} {\bibinfo
   {journal} {Nature Phys.}\ }\textbf {\bibinfo {volume} {8}},\ \bibinfo
  {pages} {277} (\bibinfo {year} {2012})}\BibitemShut {NoStop}%
\bibitem [{\citenamefont {Cirac}\ and\ \citenamefont
  {Zoller}(1995)}]{Cirac1995}%
  \BibitemOpen
  \bibfield  {author} {\bibinfo {author} {\bibfnamefont {J.~I.}\ \bibnamefont
  {Cirac}}\ and\ \bibinfo {author} {\bibfnamefont {P.}~\bibnamefont {Zoller}},\
  }\href@noop {} {\bibfield  {journal} {\bibinfo  {journal} {Phys. Rev. Lett.}\
  }\textbf {\bibinfo {volume} {74}},\ \bibinfo {pages} {4091} (\bibinfo {year}
  {1995})}\BibitemShut {NoStop}%
\bibitem [{\citenamefont {Monroe}\ \emph {et~al.}(1995)\citenamefont {Monroe},
  \citenamefont {Meekhof}, \citenamefont {King}, \citenamefont {Itano},\ and\
  \citenamefont {Wineland}}]{Monroe1995}%
  \BibitemOpen
  \bibfield  {author} {\bibinfo {author} {\bibfnamefont {C.}~\bibnamefont
  {Monroe}}, \bibinfo {author} {\bibfnamefont {D.~M.}\ \bibnamefont {Meekhof}},
  \bibinfo {author} {\bibfnamefont {B.~E.}\ \bibnamefont {King}}, \bibinfo
  {author} {\bibfnamefont {W.~M.}\ \bibnamefont {Itano}}, \ and\ \bibinfo
  {author} {\bibfnamefont {D.~J.}\ \bibnamefont {Wineland}},\ }\href {\doibase
  10.1103/PhysRevLett.75.4714} {\bibfield  {journal} {\bibinfo  {journal}
  {Phys. Rev. Lett.}\ }\textbf {\bibinfo {volume} {75}},\ \bibinfo {pages}
  {4714} (\bibinfo {year} {1995})}\BibitemShut {NoStop}%
\bibitem [{\citenamefont {Kim}\ \emph {et~al.}(2010)\citenamefont {Kim},
  \citenamefont {Chang}, \citenamefont {Korenblit}, \citenamefont {Islam},
  \citenamefont {Edwards}, \citenamefont {Freericks}, \citenamefont {Lin},
  \citenamefont {Duan},\ and\ \citenamefont {Monroe}}]{Kim2010}%
  \BibitemOpen
  \bibfield  {author} {\bibinfo {author} {\bibfnamefont {K.}~\bibnamefont
  {Kim}}, \bibinfo {author} {\bibfnamefont {M.-S.}\ \bibnamefont {Chang}},
  \bibinfo {author} {\bibfnamefont {S.}~\bibnamefont {Korenblit}}, \bibinfo
  {author} {\bibfnamefont {R.}~\bibnamefont {Islam}}, \bibinfo {author}
  {\bibfnamefont {E.~E.}\ \bibnamefont {Edwards}}, \bibinfo {author}
  {\bibfnamefont {J.~K.}\ \bibnamefont {Freericks}}, \bibinfo {author}
  {\bibfnamefont {G.-D.}\ \bibnamefont {Lin}}, \bibinfo {author} {\bibfnamefont
  {L.-M.}\ \bibnamefont {Duan}}, \ and\ \bibinfo {author} {\bibfnamefont
  {C.}~\bibnamefont {Monroe}},\ }\href {\doibase 10.1038/nature09071}
  {\bibfield  {journal} {\bibinfo  {journal} {Nature (London)}\ }\textbf
  {\bibinfo {volume} {465}},\ \bibinfo {pages} {590} (\bibinfo {year}
  {2010})}\BibitemShut {NoStop}%
\bibitem [{\citenamefont {Lanyon}\ \emph {et~al.}(2011)\citenamefont {Lanyon},
  \citenamefont {Hempel}, \citenamefont {Nigg}, \citenamefont {M\"uller},
  \citenamefont {Gerritsma}, \citenamefont {Z\"ahringer}, \citenamefont
  {Schindler}, \citenamefont {Barreiro}, \citenamefont {Rambach}, \citenamefont
  {Kirchmair}, \citenamefont {Hennrich}, \citenamefont {Zoller}, \citenamefont
  {Blatt},\ and\ \citenamefont {Roos}}]{Lanyon2011}%
  \BibitemOpen
  \bibfield  {author} {\bibinfo {author} {\bibfnamefont {B.~P.}\ \bibnamefont
  {Lanyon}}, \bibinfo {author} {\bibfnamefont {C.}~\bibnamefont {Hempel}},
  \bibinfo {author} {\bibfnamefont {D.}~\bibnamefont {Nigg}}, \bibinfo {author}
  {\bibfnamefont {M.}~\bibnamefont {M\"uller}}, \bibinfo {author}
  {\bibfnamefont {R.}~\bibnamefont {Gerritsma}}, \bibinfo {author}
  {\bibfnamefont {F.}~\bibnamefont {Z\"ahringer}}, \bibinfo {author}
  {\bibfnamefont {P.}~\bibnamefont {Schindler}}, \bibinfo {author}
  {\bibfnamefont {J.~T.}\ \bibnamefont {Barreiro}}, \bibinfo {author}
  {\bibfnamefont {M.}~\bibnamefont {Rambach}}, \bibinfo {author} {\bibfnamefont
  {G.}~\bibnamefont {Kirchmair}}, \bibinfo {author} {\bibfnamefont
  {M.}~\bibnamefont {Hennrich}}, \bibinfo {author} {\bibfnamefont
  {P.}~\bibnamefont {Zoller}}, \bibinfo {author} {\bibfnamefont
  {R.}~\bibnamefont {Blatt}}, \ and\ \bibinfo {author} {\bibfnamefont {C.~F.}\
  \bibnamefont {Roos}},\ }\href {\doibase 10.1126/science.1208001} {\bibfield
  {journal} {\bibinfo  {journal} {Science}\ }\textbf {\bibinfo {volume}
  {334}},\ \bibinfo {pages} {57} (\bibinfo {year} {2011})}\BibitemShut
  {NoStop}%
\bibitem [{\citenamefont {Britton}\ \emph {et~al.}(2012)\citenamefont
  {Britton}, \citenamefont {Sawyer}, \citenamefont {Keith}, \citenamefont
  {Wang}, \citenamefont {Freericks}, \citenamefont {Uys}, \citenamefont
  {Biercuk},\ and\ \citenamefont {Bollinger}}]{Britton2012}%
  \BibitemOpen
  \bibfield  {author} {\bibinfo {author} {\bibfnamefont {J.~W.}\ \bibnamefont
  {Britton}}, \bibinfo {author} {\bibfnamefont {B.~C.}\ \bibnamefont {Sawyer}},
  \bibinfo {author} {\bibfnamefont {A.~C.}\ \bibnamefont {Keith}}, \bibinfo
  {author} {\bibfnamefont {C.-C.~J.}\ \bibnamefont {Wang}}, \bibinfo {author}
  {\bibfnamefont {J.~K.}\ \bibnamefont {Freericks}}, \bibinfo {author}
  {\bibfnamefont {H.}~\bibnamefont {Uys}}, \bibinfo {author} {\bibfnamefont
  {M.~J.}\ \bibnamefont {Biercuk}}, \ and\ \bibinfo {author} {\bibfnamefont
  {J.~J.}\ \bibnamefont {Bollinger}},\ }\href {\doibase 10.1038/nature10981}
  {\bibfield  {journal} {\bibinfo  {journal} {Nature (London)}\ }\textbf
  {\bibinfo {volume} {484}},\ \bibinfo {pages} {489} (\bibinfo {year}
  {2012})}\BibitemShut {NoStop}%
\bibitem [{\citenamefont {Schmidt}\ \emph {et~al.}(2005)\citenamefont
  {Schmidt}, \citenamefont {Rosenband}, \citenamefont {Langer}, \citenamefont
  {Itano}, \citenamefont {Bergquist},\ and\ \citenamefont
  {Wineland}}]{Schmidt2005}%
  \BibitemOpen
  \bibfield  {author} {\bibinfo {author} {\bibfnamefont {P.~O.}\ \bibnamefont
  {Schmidt}}, \bibinfo {author} {\bibfnamefont {T.}~\bibnamefont {Rosenband}},
  \bibinfo {author} {\bibfnamefont {C.}~\bibnamefont {Langer}}, \bibinfo
  {author} {\bibfnamefont {W.~M.}\ \bibnamefont {Itano}}, \bibinfo {author}
  {\bibfnamefont {J.~C.}\ \bibnamefont {Bergquist}}, \ and\ \bibinfo {author}
  {\bibfnamefont {D.~J.}\ \bibnamefont {Wineland}},\ }\href {\doibase
  10.1126/science.1114375} {\bibfield  {journal} {\bibinfo  {journal}
  {Science}\ }\textbf {\bibinfo {volume} {309}},\ \bibinfo {pages} {749}
  (\bibinfo {year} {2005})}\BibitemShut {NoStop}%
\bibitem [{\citenamefont {Wolf}\ \emph {et~al.}(2016)\citenamefont {Wolf},
  \citenamefont {Wan}, \citenamefont {Heip}, \citenamefont {Gebert},
  \citenamefont {Shi},\ and\ \citenamefont {Schmidt}}]{Wolf2016}%
  \BibitemOpen
  \bibfield  {author} {\bibinfo {author} {\bibfnamefont {F.}~\bibnamefont
  {Wolf}}, \bibinfo {author} {\bibfnamefont {Y.}~\bibnamefont {Wan}}, \bibinfo
  {author} {\bibfnamefont {J.~C.}\ \bibnamefont {Heip}}, \bibinfo {author}
  {\bibfnamefont {F.}~\bibnamefont {Gebert}}, \bibinfo {author} {\bibfnamefont
  {C.}~\bibnamefont {Shi}}, \ and\ \bibinfo {author} {\bibfnamefont {P.~O.}\
  \bibnamefont {Schmidt}},\ }\href {\doibase 10.1038/nature16513} {\bibfield
  {journal} {\bibinfo  {journal} {Nature}\ }\textbf {\bibinfo {volume} {530}},\
  \bibinfo {pages} {457} (\bibinfo {year} {2016})}\BibitemShut {NoStop}%
\bibitem [{\citenamefont {Ozeri}\ \emph {et~al.}(2007)\citenamefont {Ozeri},
  \citenamefont {Itano}, \citenamefont {Blakestad}, \citenamefont {Britton},
  \citenamefont {Chiaverini}, \citenamefont {Jost}, \citenamefont {Langer},
  \citenamefont {Leibfried}, \citenamefont {Reichle}, \citenamefont {Seidelin},
  \citenamefont {Wesenberg},\ and\ \citenamefont {Wineland}}]{Ozeri2007}%
  \BibitemOpen
  \bibfield  {author} {\bibinfo {author} {\bibfnamefont {R.}~\bibnamefont
  {Ozeri}}, \bibinfo {author} {\bibfnamefont {W.~M.}\ \bibnamefont {Itano}},
  \bibinfo {author} {\bibfnamefont {R.~B.}\ \bibnamefont {Blakestad}}, \bibinfo
  {author} {\bibfnamefont {J.}~\bibnamefont {Britton}}, \bibinfo {author}
  {\bibfnamefont {J.}~\bibnamefont {Chiaverini}}, \bibinfo {author}
  {\bibfnamefont {J.~D.}\ \bibnamefont {Jost}}, \bibinfo {author}
  {\bibfnamefont {C.}~\bibnamefont {Langer}}, \bibinfo {author} {\bibfnamefont
  {D.}~\bibnamefont {Leibfried}}, \bibinfo {author} {\bibfnamefont
  {R.}~\bibnamefont {Reichle}}, \bibinfo {author} {\bibfnamefont
  {S.}~\bibnamefont {Seidelin}}, \bibinfo {author} {\bibfnamefont {J.~H.}\
  \bibnamefont {Wesenberg}}, \ and\ \bibinfo {author} {\bibfnamefont {D.~J.}\
  \bibnamefont {Wineland}},\ }\href {\doibase 10.1103/PhysRevA.75.042329}
  {\bibfield  {journal} {\bibinfo  {journal} {Phys. Rev. A}\ }\textbf {\bibinfo
  {volume} {75}},\ \bibinfo {pages} {042329} (\bibinfo {year}
  {2007})}\BibitemShut {NoStop}%
\bibitem [{\citenamefont {Ballance}\ \emph {et~al.}(2016)\citenamefont
  {Ballance}, \citenamefont {Harty}, \citenamefont {Linke}, \citenamefont
  {Sepiol},\ and\ \citenamefont {Lucas}}]{Ballance2016}%
  \BibitemOpen
  \bibfield  {author} {\bibinfo {author} {\bibfnamefont {C.~J.}\ \bibnamefont
  {Ballance}}, \bibinfo {author} {\bibfnamefont {T.~P.}\ \bibnamefont {Harty}},
  \bibinfo {author} {\bibfnamefont {N.~M.}\ \bibnamefont {Linke}}, \bibinfo
  {author} {\bibfnamefont {M.~A.}\ \bibnamefont {Sepiol}}, \ and\ \bibinfo
  {author} {\bibfnamefont {D.~M.}\ \bibnamefont {Lucas}},\ }\href {\doibase
  10.1103/PhysRevLett.117.060504} {\bibfield  {journal} {\bibinfo  {journal}
  {Phys. Rev. Lett.}\ }\textbf {\bibinfo {volume} {117}},\ \bibinfo {pages}
  {060504} (\bibinfo {year} {2016})}\BibitemShut {NoStop}%
\bibitem [{\citenamefont {Gaebler}\ \emph {et~al.}(2016)\citenamefont
  {Gaebler}, \citenamefont {Tan}, \citenamefont {Lin}, \citenamefont {Wan},
  \citenamefont {Bowler}, \citenamefont {Keith}, \citenamefont {Glancy},
  \citenamefont {Coakley}, \citenamefont {Knill}, \citenamefont {Leibfried},\
  and\ \citenamefont {Wineland}}]{Gaebler2016}%
  \BibitemOpen
  \bibfield  {author} {\bibinfo {author} {\bibfnamefont {J.~P.}\ \bibnamefont
  {Gaebler}}, \bibinfo {author} {\bibfnamefont {T.~R.}\ \bibnamefont {Tan}},
  \bibinfo {author} {\bibfnamefont {Y.}~\bibnamefont {Lin}}, \bibinfo {author}
  {\bibfnamefont {Y.}~\bibnamefont {Wan}}, \bibinfo {author} {\bibfnamefont
  {R.}~\bibnamefont {Bowler}}, \bibinfo {author} {\bibfnamefont {A.~C.}\
  \bibnamefont {Keith}}, \bibinfo {author} {\bibfnamefont {S.}~\bibnamefont
  {Glancy}}, \bibinfo {author} {\bibfnamefont {K.}~\bibnamefont {Coakley}},
  \bibinfo {author} {\bibfnamefont {E.}~\bibnamefont {Knill}}, \bibinfo
  {author} {\bibfnamefont {D.}~\bibnamefont {Leibfried}}, \ and\ \bibinfo
  {author} {\bibfnamefont {D.~J.}\ \bibnamefont {Wineland}},\ }\href {\doibase
  10.1103/PhysRevLett.117.060505} {\bibfield  {journal} {\bibinfo  {journal}
  {Phys. Rev. Lett.}\ }\textbf {\bibinfo {volume} {117}},\ \bibinfo {pages}
  {060505} (\bibinfo {year} {2016})}\BibitemShut {NoStop}%
\bibitem [{\citenamefont {Seidelin}\ \emph {et~al.}(2006)\citenamefont
  {Seidelin}, \citenamefont {Chiaverini}, \citenamefont {Reichle},
  \citenamefont {Bollinger}, \citenamefont {Leibfried}, \citenamefont
  {Britton}, \citenamefont {Wesenberg}, \citenamefont {Blakestad},
  \citenamefont {Epstein}, \citenamefont {Hume}, \citenamefont {Itano},
  \citenamefont {Jost}, \citenamefont {Langer}, \citenamefont {Ozeri},
  \citenamefont {Shiga},\ and\ \citenamefont {Wineland}}]{Seidelin2006}%
  \BibitemOpen
  \bibfield  {author} {\bibinfo {author} {\bibfnamefont {S.}~\bibnamefont
  {Seidelin}}, \bibinfo {author} {\bibfnamefont {J.}~\bibnamefont
  {Chiaverini}}, \bibinfo {author} {\bibfnamefont {R.}~\bibnamefont {Reichle}},
  \bibinfo {author} {\bibfnamefont {J.~J.}\ \bibnamefont {Bollinger}}, \bibinfo
  {author} {\bibfnamefont {D.}~\bibnamefont {Leibfried}}, \bibinfo {author}
  {\bibfnamefont {J.}~\bibnamefont {Britton}}, \bibinfo {author} {\bibfnamefont
  {J.~H.}\ \bibnamefont {Wesenberg}}, \bibinfo {author} {\bibfnamefont {R.~B.}\
  \bibnamefont {Blakestad}}, \bibinfo {author} {\bibfnamefont {R.~J.}\
  \bibnamefont {Epstein}}, \bibinfo {author} {\bibfnamefont {D.~B.}\
  \bibnamefont {Hume}}, \bibinfo {author} {\bibfnamefont {W.~M.}\ \bibnamefont
  {Itano}}, \bibinfo {author} {\bibfnamefont {J.~D.}\ \bibnamefont {Jost}},
  \bibinfo {author} {\bibfnamefont {C.}~\bibnamefont {Langer}}, \bibinfo
  {author} {\bibfnamefont {R.}~\bibnamefont {Ozeri}}, \bibinfo {author}
  {\bibfnamefont {N.}~\bibnamefont {Shiga}}, \ and\ \bibinfo {author}
  {\bibfnamefont {D.~J.}\ \bibnamefont {Wineland}},\ }\href {\doibase
  10.1103/PhysRevLett.96.253003} {\bibfield  {journal} {\bibinfo  {journal}
  {Phys. Rev. Lett.}\ }\textbf {\bibinfo {volume} {96}},\ \bibinfo {pages}
  {253003} (\bibinfo {year} {2006})}\BibitemShut {NoStop}%
\bibitem [{\citenamefont {Brown}\ \emph {et~al.}(2011)\citenamefont {Brown},
  \citenamefont {Wilson}, \citenamefont {Colombe}, \citenamefont {Ospelkaus},
  \citenamefont {Meier}, \citenamefont {Knill}, \citenamefont {Leibfried},\
  and\ \citenamefont {Wineland}}]{Brown2011}%
  \BibitemOpen
  \bibfield  {author} {\bibinfo {author} {\bibfnamefont {K.~R.}\ \bibnamefont
  {Brown}}, \bibinfo {author} {\bibfnamefont {A.~C.}\ \bibnamefont {Wilson}},
  \bibinfo {author} {\bibfnamefont {Y.}~\bibnamefont {Colombe}}, \bibinfo
  {author} {\bibfnamefont {C.}~\bibnamefont {Ospelkaus}}, \bibinfo {author}
  {\bibfnamefont {A.~M.}\ \bibnamefont {Meier}}, \bibinfo {author}
  {\bibfnamefont {E.}~\bibnamefont {Knill}}, \bibinfo {author} {\bibfnamefont
  {D.}~\bibnamefont {Leibfried}}, \ and\ \bibinfo {author} {\bibfnamefont
  {D.~J.}\ \bibnamefont {Wineland}},\ }\href {\doibase
  doi.org/10.1103/PhysRevA.84.030303} {\bibfield  {journal} {\bibinfo
  {journal} {Phys. Rev. A}\ }\textbf {\bibinfo {volume} {84}},\ \bibinfo
  {pages} {030303} (\bibinfo {year} {2011})}\BibitemShut {NoStop}%
\bibitem [{\citenamefont {Harty}\ \emph {et~al.}(2014)\citenamefont {Harty},
  \citenamefont {Allcock}, \citenamefont {Ballance}, \citenamefont {Guidoni},
  \citenamefont {Janacek}, \citenamefont {Linke}, \citenamefont {Stacey},\ and\
  \citenamefont {Lucas}}]{Harty2014}%
  \BibitemOpen
  \bibfield  {author} {\bibinfo {author} {\bibfnamefont {T.~P.}\ \bibnamefont
  {Harty}}, \bibinfo {author} {\bibfnamefont {D.~T.~C.}\ \bibnamefont
  {Allcock}}, \bibinfo {author} {\bibfnamefont {C.~J.}\ \bibnamefont
  {Ballance}}, \bibinfo {author} {\bibfnamefont {L.}~\bibnamefont {Guidoni}},
  \bibinfo {author} {\bibfnamefont {H.~A.}\ \bibnamefont {Janacek}}, \bibinfo
  {author} {\bibfnamefont {N.~M.}\ \bibnamefont {Linke}}, \bibinfo {author}
  {\bibfnamefont {D.~N.}\ \bibnamefont {Stacey}}, \ and\ \bibinfo {author}
  {\bibfnamefont {D.~M.}\ \bibnamefont {Lucas}},\ }\href {\doibase
  10.1103/PhysRevLett.113.220501} {\bibfield  {journal} {\bibinfo  {journal}
  {Phys. Rev. Lett.}\ }\textbf {\bibinfo {volume} {113}},\ \bibinfo {pages}
  {220501} (\bibinfo {year} {2014})}\BibitemShut {NoStop}%
\bibitem [{\citenamefont {Warring}\ \emph
  {et~al.}(2013{\natexlab{a}})\citenamefont {Warring}, \citenamefont
  {Ospelkaus}, \citenamefont {Colombe}, \citenamefont {J{\"o}rdens},
  \citenamefont {Leibfried},\ and\ \citenamefont {Wineland}}]{Warring2013a}%
  \BibitemOpen
  \bibfield  {author} {\bibinfo {author} {\bibfnamefont {U.}~\bibnamefont
  {Warring}}, \bibinfo {author} {\bibfnamefont {C.}~\bibnamefont {Ospelkaus}},
  \bibinfo {author} {\bibfnamefont {Y.}~\bibnamefont {Colombe}}, \bibinfo
  {author} {\bibfnamefont {R.}~\bibnamefont {J{\"o}rdens}}, \bibinfo {author}
  {\bibfnamefont {D.}~\bibnamefont {Leibfried}}, \ and\ \bibinfo {author}
  {\bibfnamefont {D.~J.}\ \bibnamefont {Wineland}},\ }\href {\doibase
  10.1103/PhysRevLett.110.173002} {\bibfield  {journal} {\bibinfo  {journal}
  {Phys. Rev. Lett.}\ }\textbf {\bibinfo {volume} {110}},\ \bibinfo {pages}
  {173002} (\bibinfo {year} {2013}{\natexlab{a}})}\BibitemShut {NoStop}%
\bibitem [{\citenamefont {Aude~Craik}\ \emph {et~al.}(2014)\citenamefont
  {Aude~Craik}, \citenamefont {Linke}, \citenamefont {Harty}, \citenamefont
  {Ballance}, \citenamefont {Lucas}, \citenamefont {Steane},\ and\
  \citenamefont {Allcock}}]{Craik2014}%
  \BibitemOpen
  \bibfield  {author} {\bibinfo {author} {\bibfnamefont {D.~P.~L.}\
  \bibnamefont {Aude~Craik}}, \bibinfo {author} {\bibfnamefont {N.~M.}\
  \bibnamefont {Linke}}, \bibinfo {author} {\bibfnamefont {T.~P.}\ \bibnamefont
  {Harty}}, \bibinfo {author} {\bibfnamefont {C.~J.}\ \bibnamefont {Ballance}},
  \bibinfo {author} {\bibfnamefont {D.~M.}\ \bibnamefont {Lucas}}, \bibinfo
  {author} {\bibfnamefont {A.~M.}\ \bibnamefont {Steane}}, \ and\ \bibinfo
  {author} {\bibfnamefont {D.~T.~C.}\ \bibnamefont {Allcock}},\ }\href
  {\doibase 10.1007/s00340-013-5716-7} {\bibfield  {journal} {\bibinfo
  {journal} {Appl. Phys. B}\ }\textbf {\bibinfo {volume} {114}},\ \bibinfo
  {pages} {3} (\bibinfo {year} {2014})}\BibitemShut {NoStop}%
\bibitem [{\citenamefont {Mintert}\ and\ \citenamefont
  {Wunderlich}(2001)}]{Mintert2001}%
  \BibitemOpen
  \bibfield  {author} {\bibinfo {author} {\bibfnamefont {F.}~\bibnamefont
  {Mintert}}\ and\ \bibinfo {author} {\bibfnamefont {C.}~\bibnamefont
  {Wunderlich}},\ }\href {\doibase 10.1103/PhysRevLett.87.257904} {\bibfield
  {journal} {\bibinfo  {journal} {Phys. Rev. Lett.}\ }\textbf {\bibinfo
  {volume} {87}},\ \bibinfo {pages} {257904} (\bibinfo {year}
  {2001})}\BibitemShut {NoStop}%
\bibitem [{\citenamefont {Johanning}\ \emph {et~al.}(2009)\citenamefont
  {Johanning}, \citenamefont {Braun}, \citenamefont {Timoney}, \citenamefont
  {Elman}, \citenamefont {Neuhauser},\ and\ \citenamefont
  {Wunderlich}}]{Johanning2009}%
  \BibitemOpen
  \bibfield  {author} {\bibinfo {author} {\bibfnamefont {M.}~\bibnamefont
  {Johanning}}, \bibinfo {author} {\bibfnamefont {A.}~\bibnamefont {Braun}},
  \bibinfo {author} {\bibfnamefont {N.}~\bibnamefont {Timoney}}, \bibinfo
  {author} {\bibfnamefont {V.}~\bibnamefont {Elman}}, \bibinfo {author}
  {\bibfnamefont {W.}~\bibnamefont {Neuhauser}}, \ and\ \bibinfo {author}
  {\bibfnamefont {C.}~\bibnamefont {Wunderlich}},\ }\href {\doibase
  10.1103/PhysRevLett.102.073004} {\bibfield  {journal} {\bibinfo  {journal}
  {Phys. Rev. Lett.}\ }\textbf {\bibinfo {volume} {102}},\ \bibinfo {pages}
  {073004} (\bibinfo {year} {2009})}\BibitemShut {NoStop}%
\bibitem [{\citenamefont {Khromova}\ \emph {et~al.}(2012)\citenamefont
  {Khromova}, \citenamefont {Piltz}, \citenamefont {Scharfenberger},
  \citenamefont {Gloger}, \citenamefont {Johanning}, \citenamefont {Var\'on},\
  and\ \citenamefont {Wunderlich}}]{Khromova2012}%
  \BibitemOpen
  \bibfield  {author} {\bibinfo {author} {\bibfnamefont {A.}~\bibnamefont
  {Khromova}}, \bibinfo {author} {\bibfnamefont {C.}~\bibnamefont {Piltz}},
  \bibinfo {author} {\bibfnamefont {B.}~\bibnamefont {Scharfenberger}},
  \bibinfo {author} {\bibfnamefont {T.~F.}\ \bibnamefont {Gloger}}, \bibinfo
  {author} {\bibfnamefont {M.}~\bibnamefont {Johanning}}, \bibinfo {author}
  {\bibfnamefont {A.~F.}\ \bibnamefont {Var\'on}}, \ and\ \bibinfo {author}
  {\bibfnamefont {C.}~\bibnamefont {Wunderlich}},\ }\href {\doibase
  10.1103/PhysRevLett.108.220502} {\bibfield  {journal} {\bibinfo  {journal}
  {Phys. Rev. Lett.}\ }\textbf {\bibinfo {volume} {108}},\ \bibinfo {pages}
  {220502} (\bibinfo {year} {2012})}\BibitemShut {NoStop}%
\bibitem [{\citenamefont {Ospelkaus}\ \emph {et~al.}(2008)\citenamefont
  {Ospelkaus}, \citenamefont {Langer}, \citenamefont {Amini}, \citenamefont
  {Brown}, \citenamefont {Leibfried},\ and\ \citenamefont
  {Wineland}}]{Ospelkaus2008}%
  \BibitemOpen
  \bibfield  {author} {\bibinfo {author} {\bibfnamefont {C.}~\bibnamefont
  {Ospelkaus}}, \bibinfo {author} {\bibfnamefont {C.~E.}\ \bibnamefont
  {Langer}}, \bibinfo {author} {\bibfnamefont {J.~M.}\ \bibnamefont {Amini}},
  \bibinfo {author} {\bibfnamefont {K.~R.}\ \bibnamefont {Brown}}, \bibinfo
  {author} {\bibfnamefont {D.}~\bibnamefont {Leibfried}}, \ and\ \bibinfo
  {author} {\bibfnamefont {D.~J.}\ \bibnamefont {Wineland}},\ }\href {\doibase
  10.1103/PhysRevLett.101.090502} {\bibfield  {journal} {\bibinfo  {journal}
  {Phys. Rev. Lett.}\ }\textbf {\bibinfo {volume} {101}},\ \bibinfo {pages}
  {090502} (\bibinfo {year} {2008})}\BibitemShut {NoStop}%
\bibitem [{\citenamefont {Ospelkaus}\ \emph {et~al.}(2011)\citenamefont
  {Ospelkaus}, \citenamefont {Warring}, \citenamefont {Colombe}, \citenamefont
  {Brown}, \citenamefont {Amini}, \citenamefont {Leibfried},\ and\
  \citenamefont {Wineland}}]{Ospelkaus2011}%
  \BibitemOpen
  \bibfield  {author} {\bibinfo {author} {\bibfnamefont {C.}~\bibnamefont
  {Ospelkaus}}, \bibinfo {author} {\bibfnamefont {U.}~\bibnamefont {Warring}},
  \bibinfo {author} {\bibfnamefont {Y.}~\bibnamefont {Colombe}}, \bibinfo
  {author} {\bibfnamefont {K.}~\bibnamefont {Brown}}, \bibinfo {author}
  {\bibfnamefont {J.~M.}\ \bibnamefont {Amini}}, \bibinfo {author}
  {\bibfnamefont {D.}~\bibnamefont {Leibfried}}, \ and\ \bibinfo {author}
  {\bibfnamefont {D.~J.}\ \bibnamefont {Wineland}},\ }\href {\doibase
  10.1038/nature10290} {\bibfield  {journal} {\bibinfo  {journal} {Nature
  (London)}\ }\textbf {\bibinfo {volume} {476}},\ \bibinfo {pages} {181}
  (\bibinfo {year} {2011})}\BibitemShut {NoStop}%
\bibitem [{\citenamefont {Weidt}\ \emph {et~al.}(2016)\citenamefont {Weidt},
  \citenamefont {Randall}, \citenamefont {Webster}, \citenamefont {Lake},
  \citenamefont {Webb}, \citenamefont {Cohen}, \citenamefont {Navickas},
  \citenamefont {Lekitsch}, \citenamefont {Retzker},\ and\ \citenamefont
  {Hensinger}}]{Weidt2016}%
  \BibitemOpen
  \bibfield  {author} {\bibinfo {author} {\bibfnamefont {S.}~\bibnamefont
  {Weidt}}, \bibinfo {author} {\bibfnamefont {J.}~\bibnamefont {Randall}},
  \bibinfo {author} {\bibfnamefont {S.~C.}\ \bibnamefont {Webster}}, \bibinfo
  {author} {\bibfnamefont {K.}~\bibnamefont {Lake}}, \bibinfo {author}
  {\bibfnamefont {A.~E.}\ \bibnamefont {Webb}}, \bibinfo {author}
  {\bibfnamefont {I.}~\bibnamefont {Cohen}}, \bibinfo {author} {\bibfnamefont
  {T.}~\bibnamefont {Navickas}}, \bibinfo {author} {\bibfnamefont
  {B.}~\bibnamefont {Lekitsch}}, \bibinfo {author} {\bibfnamefont
  {A.}~\bibnamefont {Retzker}}, \ and\ \bibinfo {author} {\bibfnamefont
  {W.~K.}\ \bibnamefont {Hensinger}},\ }\href {\doibase
  10.1103/PhysRevLett.117.220501} {\bibfield  {journal} {\bibinfo  {journal}
  {Phys. Rev. Lett.}\ }\textbf {\bibinfo {volume} {117}},\ \bibinfo {pages}
  {220501} (\bibinfo {year} {2016})}\BibitemShut {NoStop}%
\bibitem [{\citenamefont {Harty}\ \emph {et~al.}(2016)\citenamefont {Harty},
  \citenamefont {Sepiol}, \citenamefont {Allcock}, \citenamefont {Ballance},
  \citenamefont {Tarlton},\ and\ \citenamefont {Lucas}}]{Harty2016}%
  \BibitemOpen
  \bibfield  {author} {\bibinfo {author} {\bibfnamefont {T.~P.}\ \bibnamefont
  {Harty}}, \bibinfo {author} {\bibfnamefont {M.~A.}\ \bibnamefont {Sepiol}},
  \bibinfo {author} {\bibfnamefont {D.~T.~C.}\ \bibnamefont {Allcock}},
  \bibinfo {author} {\bibfnamefont {C.~J.}\ \bibnamefont {Ballance}}, \bibinfo
  {author} {\bibfnamefont {J.~E.}\ \bibnamefont {Tarlton}}, \ and\ \bibinfo
  {author} {\bibfnamefont {D.~M.}\ \bibnamefont {Lucas}},\ }\href {\doibase
  10.1103/PhysRevLett.117.140501} {\bibfield  {journal} {\bibinfo  {journal}
  {Phys. Rev. Lett.}\ }\textbf {\bibinfo {volume} {117}},\ \bibinfo {pages}
  {140501} (\bibinfo {year} {2016})}\BibitemShut {NoStop}%
\bibitem [{\citenamefont {Brownnutt}\ \emph {et~al.}(2015)\citenamefont
  {Brownnutt}, \citenamefont {Kumph}, \citenamefont {Rabl},\ and\ \citenamefont
  {Blatt}}]{Brownutt2015}%
  \BibitemOpen
  \bibfield  {author} {\bibinfo {author} {\bibfnamefont {M.}~\bibnamefont
  {Brownnutt}}, \bibinfo {author} {\bibfnamefont {M.}~\bibnamefont {Kumph}},
  \bibinfo {author} {\bibfnamefont {P.}~\bibnamefont {Rabl}}, \ and\ \bibinfo
  {author} {\bibfnamefont {R.}~\bibnamefont {Blatt}},\ }\href {\doibase
  10.1103/RevModPhys.87.1419} {\bibfield  {journal} {\bibinfo  {journal} {Rev.
  Mod. Phys.}\ }\textbf {\bibinfo {volume} {87}},\ \bibinfo {pages} {1419}
  (\bibinfo {year} {2015})}\BibitemShut {NoStop}%
\bibitem [{\citenamefont {Ding}\ \emph {et~al.}(2014)\citenamefont {Ding},
  \citenamefont {Loh}, \citenamefont {Hablutzel}, \citenamefont {Gao},
  \citenamefont {Maslennikov},\ and\ \citenamefont {Matsukevich}}]{Ding2014}%
  \BibitemOpen
  \bibfield  {author} {\bibinfo {author} {\bibfnamefont {S.}~\bibnamefont
  {Ding}}, \bibinfo {author} {\bibfnamefont {H.}~\bibnamefont {Loh}}, \bibinfo
  {author} {\bibfnamefont {R.}~\bibnamefont {Hablutzel}}, \bibinfo {author}
  {\bibfnamefont {M.}~\bibnamefont {Gao}}, \bibinfo {author} {\bibfnamefont
  {G.}~\bibnamefont {Maslennikov}}, \ and\ \bibinfo {author} {\bibfnamefont
  {D.}~\bibnamefont {Matsukevich}},\ }\href {\doibase
  10.1103/PhysRevLett.113.073002} {\bibfield  {journal} {\bibinfo  {journal}
  {Phys. Rev. Lett.}\ }\textbf {\bibinfo {volume} {113}},\ \bibinfo {pages}
  {073002} (\bibinfo {year} {2014})}\BibitemShut {NoStop}%
\bibitem [{\citenamefont {F{\"o}rster}\ \emph {et~al.}(2009)\citenamefont
  {F{\"o}rster}, \citenamefont {Karski}, \citenamefont {Choi}, \citenamefont
  {Steffen}, \citenamefont {Alt}, \citenamefont {Meschede}, \citenamefont
  {Widera}, \citenamefont {Montano}, \citenamefont {Lee}, \citenamefont
  {Rakreungdet},\ and\ \citenamefont {Jessen}}]{Foerster2009}%
  \BibitemOpen
  \bibfield  {author} {\bibinfo {author} {\bibfnamefont {L.}~\bibnamefont
  {F{\"o}rster}}, \bibinfo {author} {\bibfnamefont {M.}~\bibnamefont {Karski}},
  \bibinfo {author} {\bibfnamefont {J.-M.}\ \bibnamefont {Choi}}, \bibinfo
  {author} {\bibfnamefont {A.}~\bibnamefont {Steffen}}, \bibinfo {author}
  {\bibfnamefont {W.}~\bibnamefont {Alt}}, \bibinfo {author} {\bibfnamefont
  {D.}~\bibnamefont {Meschede}}, \bibinfo {author} {\bibfnamefont
  {A.}~\bibnamefont {Widera}}, \bibinfo {author} {\bibfnamefont
  {E.}~\bibnamefont {Montano}}, \bibinfo {author} {\bibfnamefont {J.~H.}\
  \bibnamefont {Lee}}, \bibinfo {author} {\bibfnamefont {W.}~\bibnamefont
  {Rakreungdet}}, \ and\ \bibinfo {author} {\bibfnamefont {P.~S.}\ \bibnamefont
  {Jessen}},\ }\href {\doibase 10.1103/PhysRevLett.103.233001} {\bibfield
  {journal} {\bibinfo  {journal} {Phys. Rev. Lett.}\ }\textbf {\bibinfo
  {volume} {103}},\ \bibinfo {pages} {233001} (\bibinfo {year}
  {2009})}\BibitemShut {NoStop}%
\bibitem [{\citenamefont {Hu}\ \emph {et~al.}(2011)\citenamefont {Hu},
  \citenamefont {Yang}, \citenamefont {Xu}, \citenamefont {Zhou}, \citenamefont
  {Chen}, \citenamefont {Gao}, \citenamefont {Feng},\ and\ \citenamefont
  {Lee}}]{Hu2011}%
  \BibitemOpen
  \bibfield  {author} {\bibinfo {author} {\bibfnamefont {Y.~M.}\ \bibnamefont
  {Hu}}, \bibinfo {author} {\bibfnamefont {W.~L.}\ \bibnamefont {Yang}},
  \bibinfo {author} {\bibfnamefont {Y.~Y.}\ \bibnamefont {Xu}}, \bibinfo
  {author} {\bibfnamefont {F.}~\bibnamefont {Zhou}}, \bibinfo {author}
  {\bibfnamefont {L.}~\bibnamefont {Chen}}, \bibinfo {author} {\bibfnamefont
  {K.~L.}\ \bibnamefont {Gao}}, \bibinfo {author} {\bibfnamefont
  {M.}~\bibnamefont {Feng}}, \ and\ \bibinfo {author} {\bibfnamefont
  {C.}~\bibnamefont {Lee}},\ }\href@noop {} {\bibfield  {journal} {\bibinfo
  {journal} {New J. Phys.}\ }\textbf {\bibinfo {volume} {13}},\ \bibinfo
  {pages} {053037} (\bibinfo {year} {2011})}\BibitemShut {NoStop}%
\bibitem [{\citenamefont {Welzel}\ \emph {et~al.}(2011)\citenamefont {Welzel},
  \citenamefont {Bautista-Salvador}, \citenamefont {Abarbanel}, \citenamefont
  {Wineman-Fisher}, \citenamefont {Wunderlich}, \citenamefont {Folman},\ and\
  \citenamefont {Schmidt-Kaler}}]{Welzel2011}%
  \BibitemOpen
  \bibfield  {author} {\bibinfo {author} {\bibfnamefont {J.}~\bibnamefont
  {Welzel}}, \bibinfo {author} {\bibfnamefont {A.}~\bibnamefont
  {Bautista-Salvador}}, \bibinfo {author} {\bibfnamefont {C.}~\bibnamefont
  {Abarbanel}}, \bibinfo {author} {\bibfnamefont {V.}~\bibnamefont
  {Wineman-Fisher}}, \bibinfo {author} {\bibfnamefont {C.}~\bibnamefont
  {Wunderlich}}, \bibinfo {author} {\bibfnamefont {R.}~\bibnamefont {Folman}},
  \ and\ \bibinfo {author} {\bibfnamefont {F.}~\bibnamefont {Schmidt-Kaler}},\
  }\href@noop {} {\bibfield  {journal} {\bibinfo  {journal} {Eur. Phys. J. D}\
  }\textbf {\bibinfo {volume} {65}},\ \bibinfo {pages} {285} (\bibinfo {year}
  {2011})}\BibitemShut {NoStop}%
\bibitem [{sup()}]{supplementary}%
  \BibitemOpen
  \href@noop {} {}\bibinfo {note} {See Supplemental Material for detailed
  derivation of equations presented in text and simulations of magnetic fields
  and magnetic field gradients.}\BibitemShut {Stop}%
\bibitem [{\citenamefont {Meir}\ \emph {et~al.}(2018)\citenamefont {Meir},
  \citenamefont {Sikorsky}, \citenamefont {Ben-shlomi}, \citenamefont
  {Akerman}, \citenamefont {Pinkas}, \citenamefont {Dallal},\ and\
  \citenamefont {Ozeri}}]{Meir2018}%
  \BibitemOpen
  \bibfield  {author} {\bibinfo {author} {\bibfnamefont {Z.}~\bibnamefont
  {Meir}}, \bibinfo {author} {\bibfnamefont {T.}~\bibnamefont {Sikorsky}},
  \bibinfo {author} {\bibfnamefont {R.}~\bibnamefont {Ben-shlomi}}, \bibinfo
  {author} {\bibfnamefont {N.}~\bibnamefont {Akerman}}, \bibinfo {author}
  {\bibfnamefont {M.}~\bibnamefont {Pinkas}}, \bibinfo {author} {\bibfnamefont
  {Y.}~\bibnamefont {Dallal}}, \ and\ \bibinfo {author} {\bibfnamefont
  {R.}~\bibnamefont {Ozeri}},\ }\href {\doibase 10.1080/09500340.2017.1397217}
  {\bibfield  {journal} {\bibinfo  {journal} {J. Mod. Optics}\ }\textbf
  {\bibinfo {volume} {65}},\ \bibinfo {pages} {501} (\bibinfo {year}
  {2018})}\BibitemShut {NoStop}%
\bibitem [{\citenamefont {Weidt}\ \emph {et~al.}(2015)\citenamefont {Weidt},
  \citenamefont {Randall}, \citenamefont {Webster}, \citenamefont {Standing},
  \citenamefont {Rodriguez}, \citenamefont {Webb}, \citenamefont {Lekitsch},\
  and\ \citenamefont {Hensinger}}]{Weidt2015}%
  \BibitemOpen
  \bibfield  {author} {\bibinfo {author} {\bibfnamefont {S.}~\bibnamefont
  {Weidt}}, \bibinfo {author} {\bibfnamefont {J.}~\bibnamefont {Randall}},
  \bibinfo {author} {\bibfnamefont {S.~C.}\ \bibnamefont {Webster}}, \bibinfo
  {author} {\bibfnamefont {E.~D.}\ \bibnamefont {Standing}}, \bibinfo {author}
  {\bibfnamefont {A.}~\bibnamefont {Rodriguez}}, \bibinfo {author}
  {\bibfnamefont {A.~E.}\ \bibnamefont {Webb}}, \bibinfo {author}
  {\bibfnamefont {B.}~\bibnamefont {Lekitsch}}, \ and\ \bibinfo {author}
  {\bibfnamefont {W.~K.}\ \bibnamefont {Hensinger}},\ }\href {\doibase
  10.1103/PhysRevLett.115.013002} {\bibfield  {journal} {\bibinfo  {journal}
  {Phys. Rev. Lett.}\ }\textbf {\bibinfo {volume} {115}},\ \bibinfo {pages}
  {013002} (\bibinfo {year} {2015})}\BibitemShut {NoStop}%
\bibitem [{\citenamefont {Sriarunothai}\ \emph {et~al.}(2018)\citenamefont
  {Sriarunothai}, \citenamefont {Giri}, \citenamefont {W{\"o}lk},\ and\
  \citenamefont {Wunderlich}}]{Sriarunothai2018}%
  \BibitemOpen
  \bibfield  {author} {\bibinfo {author} {\bibfnamefont {T.}~\bibnamefont
  {Sriarunothai}}, \bibinfo {author} {\bibfnamefont {G.~S.}\ \bibnamefont
  {Giri}}, \bibinfo {author} {\bibfnamefont {S.}~\bibnamefont {W{\"o}lk}}, \
  and\ \bibinfo {author} {\bibfnamefont {C.}~\bibnamefont {Wunderlich}},\
  }\href {\doibase 10.1080/09500340.2017.1401137} {\bibfield  {journal}
  {\bibinfo  {journal} {J. Mod. Optics}\ }\textbf {\bibinfo {volume} {65}},\
  \bibinfo {pages} {560} (\bibinfo {year} {2018})}\BibitemShut {NoStop}%
\bibitem [{\citenamefont {Harris}(1978)}]{Harris1978}%
  \BibitemOpen
  \bibfield  {author} {\bibinfo {author} {\bibfnamefont {F.~J.}\ \bibnamefont
  {Harris}},\ }\href {\doibase 10.1109/PROC.1978.10837} {\bibfield  {journal}
  {\bibinfo  {journal} {Proc. IEEE}\ }\textbf {\bibinfo {volume} {66}},\
  \bibinfo {pages} {51} (\bibinfo {year} {1978})}\BibitemShut {NoStop}%
\bibitem [{\citenamefont {Warring}\ \emph
  {et~al.}(2013{\natexlab{b}})\citenamefont {Warring}, \citenamefont
  {Ospelkaus}, \citenamefont {Colombe}, \citenamefont {Brown}, \citenamefont
  {Amini}, \citenamefont {Carsjens}, \citenamefont {Leibfried},\ and\
  \citenamefont {Wineland}}]{Warring2013b}%
  \BibitemOpen
  \bibfield  {author} {\bibinfo {author} {\bibfnamefont {U.}~\bibnamefont
  {Warring}}, \bibinfo {author} {\bibfnamefont {C.}~\bibnamefont {Ospelkaus}},
  \bibinfo {author} {\bibfnamefont {Y.}~\bibnamefont {Colombe}}, \bibinfo
  {author} {\bibfnamefont {K.~R.}\ \bibnamefont {Brown}}, \bibinfo {author}
  {\bibfnamefont {J.~M.}\ \bibnamefont {Amini}}, \bibinfo {author}
  {\bibfnamefont {M.}~\bibnamefont {Carsjens}}, \bibinfo {author}
  {\bibfnamefont {D.}~\bibnamefont {Leibfried}}, \ and\ \bibinfo {author}
  {\bibfnamefont {D.~J.}\ \bibnamefont {Wineland}},\ }\href {\doibase
  10.1103/PhysRevA.87.013437} {\bibfield  {journal} {\bibinfo  {journal} {Phys.
  Rev. A}\ }\textbf {\bibinfo {volume} {87}},\ \bibinfo {pages} {013437}
  (\bibinfo {year} {2013}{\natexlab{b}})}\BibitemShut {NoStop}%
\bibitem [{\citenamefont {Welzel}\ \emph {et~al.}(2019)\citenamefont {Welzel},
  \citenamefont {Stopp},\ and\ \citenamefont {Schmidt-Kaler}}]{Welzel2018}%
  \BibitemOpen
  \bibfield  {author} {\bibinfo {author} {\bibfnamefont {J.}~\bibnamefont
  {Welzel}}, \bibinfo {author} {\bibfnamefont {F.}~\bibnamefont {Stopp}}, \
  and\ \bibinfo {author} {\bibfnamefont {F.}~\bibnamefont {Schmidt-Kaler}},\
  }\href@noop {} {\bibfield  {journal} {\bibinfo  {journal} {J. Phys. B: At.
  Mol. Opt. Phys.}\ }\textbf {\bibinfo {volume} {52}},\ \bibinfo {pages}
  {025301} (\bibinfo {year} {2019})}\BibitemShut {NoStop}%
\bibitem [{\citenamefont {Shapira}\ \emph {et~al.}(2018)\citenamefont
  {Shapira}, \citenamefont {Shaniv}, \citenamefont {Manovitz}, \citenamefont
  {Akerman},\ and\ \citenamefont {Ozeri}}]{Shapira2018}%
  \BibitemOpen
  \bibfield  {author} {\bibinfo {author} {\bibfnamefont {Y.}~\bibnamefont
  {Shapira}}, \bibinfo {author} {\bibfnamefont {R.}~\bibnamefont {Shaniv}},
  \bibinfo {author} {\bibfnamefont {T.}~\bibnamefont {Manovitz}}, \bibinfo
  {author} {\bibfnamefont {N.}~\bibnamefont {Akerman}}, \ and\ \bibinfo
  {author} {\bibfnamefont {R.}~\bibnamefont {Ozeri}},\ }\href {\doibase
  10.1103/PhysRevLett.121.180502} {\bibfield  {journal} {\bibinfo  {journal}
  {Phys. Rev. Lett.}\ }\textbf {\bibinfo {volume} {121}},\ \bibinfo {pages}
  {180502} (\bibinfo {year} {2018})}\BibitemShut {NoStop}%
\bibitem [{\citenamefont {Webb}\ \emph {et~al.}(2018)\citenamefont {Webb},
  \citenamefont {Webster}, \citenamefont {Collingbourne}, \citenamefont
  {Bretaud}, \citenamefont {Lawrence}, \citenamefont {Weidt}, \citenamefont
  {Mintert},\ and\ \citenamefont {Hensinger}}]{Webb2018}%
  \BibitemOpen
  \bibfield  {author} {\bibinfo {author} {\bibfnamefont {A.~E.}\ \bibnamefont
  {Webb}}, \bibinfo {author} {\bibfnamefont {S.~C.}\ \bibnamefont {Webster}},
  \bibinfo {author} {\bibfnamefont {S.}~\bibnamefont {Collingbourne}}, \bibinfo
  {author} {\bibfnamefont {D.}~\bibnamefont {Bretaud}}, \bibinfo {author}
  {\bibfnamefont {A.~M.}\ \bibnamefont {Lawrence}}, \bibinfo {author}
  {\bibfnamefont {S.}~\bibnamefont {Weidt}}, \bibinfo {author} {\bibfnamefont
  {F.}~\bibnamefont {Mintert}}, \ and\ \bibinfo {author} {\bibfnamefont
  {W.~K.}\ \bibnamefont {Hensinger}},\ }\href {\doibase
  10.1103/PhysRevLett.121.180501} {\bibfield  {journal} {\bibinfo  {journal}
  {Phys. Rev. Lett.}\ }\textbf {\bibinfo {volume} {121}},\ \bibinfo {pages}
  {180501} (\bibinfo {year} {2018})}\BibitemShut {NoStop}%
\bibitem [{\citenamefont {Tan}\ \emph {et~al.}(2015)\citenamefont {Tan},
  \citenamefont {Gaebler}, \citenamefont {Lin}, \citenamefont {Wan},
  \citenamefont {Bowler}, \citenamefont {Leibfried},\ and\ \citenamefont
  {Wineland}}]{Tan2015}%
  \BibitemOpen
  \bibfield  {author} {\bibinfo {author} {\bibfnamefont {T.~R.}\ \bibnamefont
  {Tan}}, \bibinfo {author} {\bibfnamefont {J.~P.}\ \bibnamefont {Gaebler}},
  \bibinfo {author} {\bibfnamefont {Y.}~\bibnamefont {Lin}}, \bibinfo {author}
  {\bibfnamefont {Y.}~\bibnamefont {Wan}}, \bibinfo {author} {\bibfnamefont
  {R.}~\bibnamefont {Bowler}}, \bibinfo {author} {\bibfnamefont
  {D.}~\bibnamefont {Leibfried}}, \ and\ \bibinfo {author} {\bibfnamefont
  {D.~J.}\ \bibnamefont {Wineland}},\ }\href {\doibase 10.1038/nature16186}
  {\bibfield  {journal} {\bibinfo  {journal} {Nature (London)}\ }\textbf
  {\bibinfo {volume} {528}},\ \bibinfo {pages} {380} (\bibinfo {year}
  {2015})}\BibitemShut {NoStop}%
\bibitem [{\citenamefont {Sutherland}\ \emph {et~al.}(2019)\citenamefont
  {Sutherland}, \citenamefont {Srinivas}, \citenamefont {Burd}, \citenamefont
  {Leibfried}, \citenamefont {Wilson}, \citenamefont {Wineland}, \citenamefont
  {Allcock}, \citenamefont {Slichter},\ and\ \citenamefont
  {Libby}}]{Sutherland2019}%
  \BibitemOpen
  \bibfield  {author} {\bibinfo {author} {\bibfnamefont {R.~T.}\ \bibnamefont
  {Sutherland}}, \bibinfo {author} {\bibfnamefont {R.}~\bibnamefont
  {Srinivas}}, \bibinfo {author} {\bibfnamefont {S.~C.}\ \bibnamefont {Burd}},
  \bibinfo {author} {\bibfnamefont {D.}~\bibnamefont {Leibfried}}, \bibinfo
  {author} {\bibfnamefont {A.~C.}\ \bibnamefont {Wilson}}, \bibinfo {author}
  {\bibfnamefont {D.~J.}\ \bibnamefont {Wineland}}, \bibinfo {author}
  {\bibfnamefont {D.~T.~C.}\ \bibnamefont {Allcock}}, \bibinfo {author}
  {\bibfnamefont {D.~H.}\ \bibnamefont {Slichter}}, \ and\ \bibinfo {author}
  {\bibfnamefont {S.~B.}\ \bibnamefont {Libby}},\ }\href {\doibase
  10.1088/1367-2630/ab0be5} {\bibfield  {journal} {\bibinfo  {journal} {New
  Journal of Physics}\ }\textbf {\bibinfo {volume} {21}},\ \bibinfo {pages}
  {033033} (\bibinfo {year} {2019})}\BibitemShut {NoStop}%
\end{thebibliography}%

\clearpage

\section{Supplemental Material}

\setcounter{equation}{0}
\renewcommand{\theequation}{S\arabic{equation}}

In this section, we present a derivation of the equations presented in the main text. We also provide details of the magnetic and electric field measurements.

\subsection{Near-motional oscillating gradient}

We first derive the sideband interaction for the case of an oscillating gradient. For simplicity, we only include a gradient term oscillating at $\omega_g$ and assume oscillating magnetic fields of frequency $\omega_g$ at the position of the ion are absent. The Hamiltonian is

\begin{align}
\begin{split}
\hat{H}_0(t)= &\frac{\hbar\omega_0}{2}\hat{\sigma}_z + \hbar\omega_r\hat{a}^\dagger\hat{a} \\
&+2\hbar\Omega_g\cos({\omega_gt})\hat{\sigma}_z(\hat{a}+\hat{a}^\dagger) \\
&+2\hbar\Omega_\mu\cos{\big((\omega_0+\delta)t\big)\hat{\sigma}_x}.
\end{split}
\end{align}

\noindent The first two terms correspond to the qubit and motional mode respectively. The next term corresponds to the gradient, which causes a spin-dependent displacement oscillating at $\omega_g$. The last term is due to the coupling of the qubit states by a microwave field detuned from the qubit transition $\omega_0$ by $\delta$. The spin-motion coupling strength $\Omega_g$ is defined in Eq.~(\ref{eq_omegas}) of the main text and $\Omega_\mu$ is given by

\begin{align}
\Omega_\mu = \frac{B_x}{2\hbar}\bra{\downarrow}\mu_x\ket{\uparrow},
\end{align}

\noindent where $B_x$ is the component of the oscillating microwave magnetic field perpendicular to the quantization axis and $\mu_x$ is the component of the ion's magnetic moment in the same direction. \noindent In the interaction picture with respect to $\hat{H}_0 = \frac{\hbar \omega_0}{2}\hat{\sigma}_z + \hbar\omega_g \hat{a}^\dagger \hat{a}$, i.e. the frame oscillating at $\omega_g$, and dropping the fast-rotating microwave terms, we obtain

\begin{align}
\begin{split}
\hat{H}_{rot}(t)= &\hbar\Omega_g \hat{\sigma}_z\left[(\hat{a}+\hat{a}^\dagger)+(\hat{a}e^{-2i\omega_gt}+\hat{a}^\dagger e^{2i\omega_gt})\right]
\\&+\hbar(\omega_r-\omega_g)\hat{a}^\dagger\hat{a}
\\
&+\hbar\Omega_\mu\left(\hat{\sigma}_+e^{-i\delta t}
+\hat{\sigma}_-e^{i\delta t}\right).
\end{split}
\end{align}

\noindent Except for the terms oscillating at $2\omega_g$, this interaction is now mathematically of the same form as the case of a static gradient with a modified ``motional frequency" $\omega_{r}-\omega_{g}$. Going into the interaction picture with respect to the $\hat{a}^\dagger\hat{a}$ term,

\begin{align}
\label{eq_h_int_oscillating}
\begin{split}
\hat{H}_I(t)=&\hbar\Omega_g \hat{\sigma}_z\Big[(\hat{a}e^{-i(\omega_r-\omega_g)t}+\hat{a}^\dagger e^{i(\omega_r-\omega_g)t})\\
&+(\hat{a}e^{-i(\omega_r+\omega_g)t}+\hat{a}^\dagger e^{i(\omega_r+\omega_g)t})\Big] \\
&+\hbar\Omega_\mu\left(\hat{\sigma}_+e^{-i\delta t}
+\hat{\sigma}_-e^{i\delta t}\right).
\end{split}
\end{align}

In order to transform into the interaction picture with respect to the
gradient term, we use the first term of the Magnus expansion to derive the propagator, as the commutator of the Hamiltonian with itself at different times is not zero. The second term in the expansion is simply a global phase, and thus higher order terms vanish as the commutator with this phase term is 0. The propagator is

\begin{align}
\begin{split}
\hat{U}_g^\dagger(t) = \exp\Big[&\frac{\Omega_g}{\omega_r-\omega_g}\hat{\sigma}_z\big(-\hat{a}(e^{-i(\omega_r-\omega_g) t}-1) 
\\&+ \hat{a}^\dagger (e^{i(\omega_r-\omega_g) t}-1)\big) +  \\
&\frac{\Omega_g}{\omega_r+\omega_g}\hat{\sigma}_z\big(-\hat{a}(e^{-i(\omega_r+\omega_g) t}-1)
\\&+ \hat{a}^\dagger (e^{i(\omega_r+\omega_g) t}-1)\big)\Big].
\end{split}
\end{align}

We use this propagator to derive the modified interaction Hamiltonian,

\begin{align}
\begin{split}
\hat{H}_I'(t) = &\hat{U}_g^\dagger (t)\hbar\Omega_{\mu}(\hat{\sigma}_+e^{-i\delta t} + \hat{\sigma}_-e^{i\delta t}) \hat{U}_g(t).
\end{split}
\end{align}

We use the Baker-Campbell-Hausdorff theorem, keeping only the lowest order term and ignoring higher order terms that scale as $\left[\Omega_g/(\omega_r\pm\omega_g)\right]^k$ for $k>1$.  We thus obtain $\hat{H}_I'(t)$:

\begin{align}
\begin{split}
\label{eq_full_int}
\hat{H}_I'(t) \simeq &\hbar\Omega_\mu(\hat{\sigma}_+e^{-i\delta t} + \hat{\sigma}_-e^{i\delta t}) \\
& +2\hbar\Omega_g\Omega_\mu(\hat{\sigma}_+e^{-i\delta t} - \hat{\sigma}_-e^{i\delta t})\\&\times \Big(\frac{1}{\omega_r-\omega_g}\big(\hat{a}^\dagger(e^{i(\omega_r-\omega_g)t}-1)-\hat{a} (e^{-i(\omega_r-\omega_g)t}-1)\big)\\
&+\frac{1}{\omega_r+\omega_g}\big(\hat{a}^\dagger(e^{i(\omega_r+\omega_g)t}-1)-\hat{a} (e^{-i(\omega_r+\omega_g)t}-1)\big)\Big).
\end{split}
\end{align}

\noindent The blue (red) sidebands occur at $\delta = +(-)(\omega_r-\omega_g)$ and $\delta = +(-)(\omega_r + \omega_g)$. The sideband interaction for the former case after eliminating fast-rotating terms is

\begin{align}
\hat{H}_{\text{sb}} =  \pm&\frac{2\hbar\Omega_g\Omega_\mu}{\omega_r-\omega_g}(\hat{\sigma}_\pm \hat{a}^{\dagger}+\hat{\sigma}_\mp \hat{a}),
\end{align}

\noindent with a sideband Rabi frequency

\begin{align}
\Omega_{\text{sb}}=\frac{2\Omega_g\Omega_\mu}{\omega_r-\omega_g}. 
\end{align}

\noindent The neglected fast-rotating terms correspond to a detuned carrier field and the additional sideband at $\omega_r + \omega_g$, which are detuned by $\delta$ and $2\omega_g$, respectively. Errors caused by these terms are coherent, and can be corrected, for example by ramping the gradient and microwaves on and off adiabatically.

We can compare this Rabi frequency to the static gradient scheme, which represents the limit $\omega_g=0$.  If we set $\omega_g=0$ in Eq.~(\ref{eq_full_int}), we find that the sideband Rabi frequency for a static gradient is

\begin{align}
\Omega_{\text{sb}}=\frac{4\Omega_g\Omega_\mu}{\omega_r}. 
\end{align}

Thus relative to the static gradient scheme, a suitable choice of $\omega_g$ would produce a larger $\Omega_{\text{sb}}$ for the same $\Omega_g, \Omega_\mu$, and $\omega_r$.  The additional factor of 2 in the static gradient Rabi frequency comes from the degeneracy of $\omega_r-\omega_g$ and $\omega_r+\omega_g$ when $\omega_g=0$.  The reader is reminded that $\Omega_g$ has an implicit $\omega_r^{-1/2}$ dependence for both the static and oscillating gradient cases, as defined in Eq.~(\ref{eq_omegas}) in the main text.  In addition, tuning $\omega_g$ close to $\omega_r$ such that ${\omega_r-\omega_g\lesssim\Omega_g}$ gives rise to non-negligible higher order sideband terms not shown in Eq.~(\ref{eq_full_int}). 

\subsection{Spin flip transitions}

We now include the effect of an oscillating magnetic field at $\omega_g$. We first consider only the qubit, the interaction corresponding to this oscillating magnetic field, and the microwave term in $\hat{H}_0$,

\begin{align}
\begin{split}
\hat{H_0}(t) = &\frac{\hbar\omega_0}{2}\hat{\sigma}_z + 2\hbar\Omega_z\cos({\omega_gt})\hat{\sigma}_z \\ &+2\hbar\Omega_\mu\cos{\big((\omega_0+\delta)t\big)\hat{\sigma}_x}.
\end{split}
\end{align}

\noindent The oscillating magnetic field term is characterized by $\Omega_z$ as defined in Eq.~(\ref{eq_omegas}). Transforming into the interaction picture with respect to the first two terms, we derive the propagator $\hat{U}_I(t)$,

\begin{align}
\begin{split}
\hat{U}^\dagger_I(t) &= \exp{\Big(\frac{i}{\hbar}\int_0^t\Big[ \frac{\hbar\omega_0}{2}\hat{\sigma}_z + 2\hbar\Omega_z\cos({\omega_gt'})\hat{\sigma}_z\Big] dt'\Big)} \\
&= \exp{\Big(\frac{i\omega_0 t}{2}\hat{\sigma}_z+\frac{2i\Omega_z}{\omega_g}\sin{(\omega_gt)}\hat{\sigma}_z\Big)}.
\end{split}
\end{align}

\noindent We use this propagator to obtain the interaction Hamiltonian

\begin{align}
\begin{split}
\hat{H}_I(t) = &\hbar\Omega_\mu(e^{-i\delta t}\hat{\sigma}_+\sum_{m=-\infty}^{\infty}J_m\left(\frac{4\Omega_z}
{\omega_g}\right)e^{im\omega_gt}) \\
& + \text{H.c},
\end{split}
\end{align}

\noindent where we have used the Jacobi-Anger expansion,

\begin{equation}
e^{iz\sin{\theta}} = \sum_{m=-\infty}^\infty J_m(z)e^{im\theta}.
\end{equation}

\noindent From this Hamiltonian we obtain spin flips when ${\delta = m\omega_g}$, given by the Hamiltonian

\begin{align}
\label{carrier}
\hat{H}_m &= \hbar\Omega_\mu J_m\left(\frac{4\Omega_z}
{\omega_g}\right)(\hat{\sigma}_++\hat{\sigma}_-),
\end{align}

\noindent after dropping fast-rotating terms. The Rabi frequency $\Omega_m$ of these transitions is

\begin{align}
\Omega_m = \Omega_\mu J_m\left(\frac{4\Omega_z}{\omega_g}\right).
\end{align}

The oscillating magnetic field modifies the microwave term from the previous section, resulting in the following Hamiltonian for the motional sideband of the $m$th spin flip sideband:

\begin{align}
\hat{H}_{\text{sb},m} =  \pm&\frac{2\hbar\Omega_g\Omega_\mu}{\omega_r-\omega_g}J_m\left(\frac{4\Omega_z}{\omega_g}\right)(\hat{\sigma}_\pm \hat{a}^{\dagger}+\hat{\sigma}_\mp \hat{a}).
\end{align}

It can be seen from this interaction that the highest motional sideband Rabi frequency is achieved for the $m=0$ spin flip transition when $B_g=0$.  

\subsection{ac Zeeman shift}

The sideband transitions occur when the microwave field is detuned from the qubit transition frequency by ${\delta=\pm(\omega_r-\omega_g)}$, where +(-) corresponds to the blue (red) sideband. However, the detuned microwave field will cause an ac Zeeman shift that shifts the qubit transition frequency. This shifts the detuning required for the sideband transition closer to resonance, with ${\delta = \pm(\omega_r-\omega_g) + \Delta_{ac}}$, where

\begin{align}
\Delta_{ac} = \delta\mp\sqrt{\delta^2+4\Omega_\mu^2}.
\end{align}

\noindent Solving for $\delta$, we obtain

\begin{align}
\delta = \pm\sqrt{(\omega_r-\omega_g)^2-4\Omega_\mu^2}.
\end{align}

\noindent Thus we require $2\Omega_\mu < |\omega_r-\omega_g|$. This limits $\Omega_{\text{sb}}$ to be less than or equal to $\Omega_g$ when using an oscillating gradient.

\subsection{Magnetic field simulation and measurement}

The current at $\omega_g$ in the trap electrodes is estimated using two different methods.  In one, we measure the magnetic field generated by each electrode using the ion, and then infer the current using a finite element simulation of the trap.  In the other, we use the same drive electronics but replace the trap with a small sense resistor, allowing a direct measurement of the current.  The methods agree at the $\sim10\,\%$ level.

\subsection{Electric field simulation and measurement}

We measure the electric field at $\omega_g$ using a technique described in Ref.~\cite{Warring2013b}, where it was used to compensate for excess micromotion. If the ion experiences an oscillating electric field at $\omega_g$, a microwave magnetic field gradient will cause spin-flip transitions when detuned from the qubit frequency by $\pm\omega_g$. The Rabi frequency of these spin flips is proportional to the amplitude of the electric field at the ion. We note that an oscillating magnetic field $B_g$ at $\omega_g$ can also give rise to spin flips when a microwave magnetic field with detuning $\omega_g$ from the qubit frequency is applied, as described in Eq.~(\ref{carrier}) with $m=1$.  To isolate the contribution to the Rabi frequency due to electric fields at $\omega_g$, we perform the measurement using the hyperfine transition ($\ket{F=3,m_F=1}\leftrightarrow\ket{F=2, m_F=1}$) which is first-order insensitive to magnetic field fluctuations at our applied magnetic field of $|\vec{B}_0|=21.3$\,mT.  This means $\Omega_z\approx0$ even if there is an oscillating magnetic field $B_g$ at $\omega_g$ at the ion, and as a result the effect in Eq.~(\ref{carrier}) will not cause spin flips. The Rabi frequency of the spin-flip transitions due to the electric field is

\begin{align}
\Omega_{\text{spin-flip}} =  \frac{2\Omega_{\mu \text{sb}} \Omega_e \omega_r}{ (\omega_r^2-\omega_g^2)}, 
\end{align}

\noindent where $\Omega_e = qEr_0/2\hbar$ and $\Omega_{\mu \text{sb}}$ is the Rabi frequency of the sideband from the oscillating gradient close to the qubit frequency. The amplitude of the oscillating electric field at the ion is $E$ and $q$ is the elementary charge. We estimate the electric field to be $\approx10\,$V/m when the magnetic field is nulled at the ion.

\end{document}